\documentclass[12pt]{article}
\pdfoutput=1
\usepackage{epsfig,graphicx,psfrag}
\usepackage{amsmath,amsfonts,stmaryrd,latexsym,amssymb}
\usepackage{url,hyperref,cite,relsize}
\usepackage{color}
\usepackage{slashed}
\usepackage{multirow,bigdelim}
\numberwithin{equation}{section}
\newcommand{\be}{\begin{equation}}
\newcommand{\ee}{\end{equation}}
\newcommand{\bea}{\begin{eqnarray}}
\newcommand{\eea}{\end{eqnarray}}
\newcommand{\bear}{\begin{eqnarray}}
\newcommand{\eear}{\end{eqnarray}}

\newcommand{\ba}{\begin{array}}
\newcommand{\ea}{\end{array}}

\newcommand{\bpm}{\begin{pmatrix}}
\newcommand{\epm}{\end{pmatrix}}

\newcommand{\AbFB}{A^{0,b}_{\rm{FB}}}
\newcommand{\asexp}{\alpha_S(M^2_Z)_{\rm{avg.}}}

\newcommand{\Ab}{\mathcal{A}_b}

\newcommand{\AlFB}{A^{0,l}_{\rm{FB}}}
\newcommand{\Al}{\mathcal{A}_l}
\newcommand{\Ae}{\mathcal{A}_e}

\newcommand{\dgLb}{\delta g_{Lb}}
\newcommand{\dgRb}{\delta g_{Rb}}

\def\eg{{\it e.g.}}

\newcommand{\hl}[1]{{\bf\color{blue}#1}}

\begin{document}

\baselineskip=18pt \pagestyle{plain} \setcounter{page}{1}

\vspace*{1.5cm}

\begin{center}

{\Large \bf The \boldmath $Zb\bar{b}$ Couplings at Future \boldmath $e^+e^-$ Colliders} \\ [9mm]

{\normalsize \bf Stefania Gori,$^{a,}$\footnote{~sgori@perimeterinstitute.ca} ~  Jiayin Gu,$^{b,}$\footnote{~gujy@ihep.ac.cn} ~  Lian-Tao Wang$\, ^{c,d,}$\footnote{~liantaow@uchicago.edu} \\ [4mm]
{\small {\it
$^a$ Perimeter Institute for Theoretical Physics, \\ 31 Caroline St. N, Waterloo, Ontario, Canada N2L 2Y5. \\ [2mm]
$^b$ Center for Future High Energy Physics, Institute of High Energy Physics, \\ Chinese Academy of Sciences, Beijing 100049, China. \\ [2mm]
$^c$ Enrico Fermi Institute, University of Chicago, Chicago, IL 60637.\\ [2mm]
$^d$ Kavli Institute for Cosmological Physics, University of Chicago, Chicago, IL 60637.
}}\\
}

%\date{ } \maketitle
\end{center}

\vspace*{0.2cm}

\begin{abstract}
Many new physics models predict sizable modifications to the SM $Zb\bar{b}$ couplings, while the corresponding measurements at LEP and SLC exhibit some discrepancy with the SM predictions.  After updating the current results on the $Zb\bar{b}$ coupling constraints from global fits, we list the observables that are most important for improving the $Zb\bar{b}$ coupling constraints and estimate the expected precision reach of three proposed future $e^+e^-$ colliders, CEPC, ILC and FCC-ee.  We consider both the case that the results are SM-like and the one that the $Zb\bar{b}$ couplings deviate significantly from the SM predictions.  We show that, if we assume the value of the $Zb\bar{b}$ couplings to be within $68\%$~CL of the current measurements, any one of the three colliders will be able to rule out the SM with more than $99.9999\%$~CL ($5\sigma$).  We study the implications of the improved $Zb\bar{b}$ coupling constraints on new physics models, and point out their complementarity with the constraints from the direct search of new physics particles at the LHC, as well as with Higgs precision measurements.  Our results provide a further motivation for the construction of future $e^+e^-$ colliders.
\end{abstract}

\newpage
{%\small 
\tableofcontents}
\newpage

\section{Introduction}
\label{sec:intro}

The LHC has just started to run at an unprecedented center-of-mass energy, 13~TeV, and will be able to probe new physics (NP) at higher energies.  At the same time, precision measurements of electroweak physics at future $e^+e^-$ machines will also offer powerful probes of Beyond the Standard Model (BSM) physics.
 
The next lepton collider may not be too distant in the future from us.  Several compelling plans exist, including the International Linear Collider (ILC) \cite{Baer:2013cma}, FCC-ee, formerly known as TLEP \cite{Gomez-Ceballos:2013zzn}, and the Circular Electron-Position Collider (CEPC) \cite{CEPCPreCDR}.  With the discovery of the Higgs boson, the primary goal of such future $e^+e^-$ colliders will be to produce a large sample of Higgs boson events at around $\sim250$~GeV to precisely measure the Higgs boson's properties, acting as a ``Higgs factory".  On the other hand, such $e^+e^-$ colliders could also collect large amount of data around the $Z$-pole, producing several orders of magnitude more $Z$-bosons than what was produced at LEP.  The large amount of $Z$-pole data would greatly improve the measurement of the several electroweak precision observables (EWPOs), which could provide strong constraints on NP.  

Several studies on the measurement of EWPOs at future $e^+e^-$ colliders have been performed \cite{Baak:2013fwa,Fan:2014vta, Fan:2014axa,Curtin:2014cca}, 
mainly focusing on the oblique corrections parameterized by the Peskin-Takeuchi parameters $S$ and $T$\cite{Peskin:1991sw}.  NP could also have sizable non-universal corrections.  The corrections to the $Zb\bar{b}$ vertex is particularly interesting, and quite generic in NP models.  For example, being the left-handed top and bottom quarks in the same electro-weak (EW) doublet, new physics that couples to the top quark usually also affects the $Zb\bar{b}$ couplings\cite{Peccei:1990uv}.  Composite Higgs models with light top partners usually predict a large correction to the $Zb_L\bar{b}_L$ coupling, unless it is protected by some symmetry analogous to the custodial symmetry that protects the weak isospin \cite{Agashe:2006at}.  Additionally, heavy Higgs bosons typically couple more strongly to the third generation quarks and modify the $Zb\bar{b}$ coupling through loops\cite{Haber:1999zh}.  

The story is even more interesting on the experimental side.  At LEP, the left and right handed $Zb\bar{b}$ couplings are mainly determined by two measurements at the $Z$-pole: $R^0_b$, the ratio of the $Z\to b\bar{b}$ partial width to the inclusive hadronic width, and $\AbFB$, the forward-backward asymmetry of the bottom quark.  The measured value of $R^0_b$ agrees with the most recent two-loop calculation of its SM prediction within $1\, \sigma$ \cite{ALEPH:2005ab, Freitas:2012sy, Baak:2014ora}. $\AbFB$, instead, exhibits a long-standing discrepancy with the SM prediction with a significance at around $2.5\,\sigma$ \cite{ALEPH:2005ab, Baak:2014ora}.  In addition, SLD directly measured the bottom quark asymmetry with longitudinal beam polarizations, $\mathcal{A}_b$, which is consistent with the SM prediction within $1\, \sigma$ but slightly prefers a shift in the same direction as the LEP $\AbFB$ measurement does.  The measured values and SM predictions of $R^0_b$, $\AbFB$ and $\mathcal{A}_b$ are summarized in Table~\ref{tab:ra0}\footnote{ Ref.~\cite{Freitas:2014hra} quotes a value of $0.00015$ as the theoretical uncertainty for the SM prediction for $R^0_b$.  Either way, the current theoretical uncertainty of $R^0_b$ is too small to have an impact on the current precision data.}.  To obtain the desired modification for $\AbFB$ without violating the experimental constraint on $R^0_b$ and $\mathcal{A}_b$, a simultaneous modification of both the $Zb_L\bar{b}_L$ and $Zb_R\bar{b}_R$ couplings is required.  To obtain the best estimation for the preferred values of the $Zb_L\bar{b}_L$ and $Zb_R\bar{b}_R$ couplings, a global fit to all precision data has to be performed (see e.g. \cite{Batell:2012ca, Ciuchini:2013pca}, for earlier studies).

\begin{table}
\centering
\begin{tabular}{c|cc}
&  measured value & SM prediction \\ \hline
$R^0_b$  & $0.21629\pm0.00066$   &  $0.21578 \pm 0.00011$   \\
$\AbFB$  & $0.0992\pm0.0016$   &  $0.1032 \pm 0.0004$  \\
$\mathcal{A}_b$  & $0.923\pm0.020$    & $0.93463\pm 0.00004$     \\
\end{tabular}
\caption{The measured values and SM predictions of $R^0_b$, $\AbFB$ and $\mathcal{A}_b$ according to the most recent result from the Gfitter group \cite{Baak:2014ora}.}
\label{tab:ra0}
\end{table}

A future $e^+e^-$ collider offers great opportunities for further studies on the $Zb\bar{b}$ couplings.  With the huge improvement on statistics at $Z$-pole, it will surely have the potential to resolve the $\AbFB$ discrepancy at LEP.  If the result agrees with the SM predictions, a $e^+e^-$ collider can provide very strong constraint on NP models; if the LEP $\AbFB$ discrepancy does come from NP, a $e^+e^-$ collider will have the potential to rule out the SM with enough significance, therefore providing strong indirect evidence for physics beyond the SM.  In either case, the results would greatly improve our understanding of fundamental particle physics. 
In this paper, we perform a study of the constraints on non-universal modifications of the $Zb\bar{b}$ couplings from prospective precisions at future $e^+e^-$ colliders, which, to our best knowledge, is the first study of such kind. 

The rest of this paper is organized as follows.  In Section~\ref{sec:current}, we review the current constraints on the $Zb\bar{b}$ couplings and discuss the importance of including the strong coupling constant in the global fit. In Section~\ref{sec:future}, we compare the precision reaches of the three proposed $e^+e^-$ colliders and outline the most important measurements needed for improving the $Zb\bar{b}$ coupling constraints.  We then perform a model independent analysis to constrain the effective Lagrangian responsible of the modifications of the $Zb\bar b$ vertex, in both the case that the results are SM-like and the one that NP causes a significant deviation in the bottom asymmetries. In Section~\ref{sec:model}, we interpret these constraints on specific NP scenarios including two Higgs doublet models (2HDMs), composite Higgs models and the Beautiful Mirror Model.  We will compare these constraints to the direct reach of the NP particles at the LHC, as well as to the constraints from oblique parameters and Higgs precision measurements. In Section~\ref{sec:con}, we present our conclusions. Finally, in Appendix~\ref{app:theo}, we discuss our treatment of theory uncertainties.

%%%%%%%%%%%%%%%%%%%%%%%%%%%%%%%%%%%%%%%%%%%%%%%%%%%%%%%%
%%%%%%%%%%%%%%%%%%%%%%%%%%%%%%%%%%%%%%%%%%%%%%%%%%%%%%%%

\section{Current constraints on the \boldmath $Zb\bar{b}$ couplings}
\label{sec:current}

In this section we present the constraints on the $Zb\bar{b}$ couplings from the global fit of the current precision electroweak data.  We follow closely the fit procedure of the Gfitter group\cite{Flacher:2008zq, Baak:2014ora}\footnote{For a global fit in the context of the Standard Model effective field theory, see e.g.~\cite{Pomarol:2013zra,Ellis:2014jta,Falkowski:2014tna,Berthier:2015oma,Berthier:2015gja}.}, with the $Z$-pole data from LEP, SLD\cite{ALEPH:2005ab} along with the measurements of the $W$, top and Higgs masses\cite{ Agashe:2014kda, ATLAS:2014wva, Aad:2014aba, CMS:2014ega} and the hadronic contribution to the running fine structure constant\cite{Davier:2010nc}\footnote{It should be noted that there also exist non-trivial bounds on the $Zb\bar{b}$ couplings from hadron colliders, as pointed out in \eg ~Ref.~\cite{Murphy:2015cha}, although the precision can not compete with the one obtained from lepton colliders.}. In particular, our procedure uses the input parameters and the electroweak and QCD corrections to the electroweak observables of the latest GFitter analysis \cite{Baak:2014ora}, even though it does not make use of the GFitter public code. 

There are, in fact, a few important differences between our procedure and the one of the Gfitter group.  First, we include the world average of the strong coupling constant $\alpha_S(M^2_Z)$\cite{Pich:2013sqa, Agashe:2014kda} (denoted as $\asexp$) as a constraint in the fit\footnote{The value we use is $ \Delta\alpha^{(5)}_{\rm had}=(2757\pm 10)\times 10^{-5}$.}.  Later in this section, we will show that the inclusion of $\asexp$ has a moderate, but non-negligible, impact on the constraints of the $Zb\bar{b}$ couplings\footnote{Ref.~\cite{Ciuchini:2013pca}, \cite{Ciuchini:2014dea} and \cite{Agashe:2014kda} included $\asexp$ but did not comment on its impact on the $Zb\bar{b}$ couplings.}.  Second, the observable $\sin^2{\theta^l_{\rm{eff}}} (Q_{\rm FB})$ is not included in our fit. $\sin^2{\theta^l_{\rm{eff}}} (Q_{\rm FB})$ is a direct measurement of the leptonic effective weak mixing angle at LEP, using the charge forward backward asymmetry.  Its measurement has a strong model dependence and the LEP result explicitly assumes the SM, therefore it is difficult to interpret this measurement in the presence of vertex corrections.  In practice, this observable is not precisely measured and has a small impact in the global fit.  To summarize, the SM free parameters considered in our fit are
$M_H \, , ~M_Z \,, m_{\rm top} \,,   \Delta\alpha^{(5)}_{\rm had}$ and $\alpha_S(M^2_Z)$. The additional observables included in our fit are $\Gamma_Z$, 
$\sigma^0_{\rm{had}}$, $R^0_l$, $A^{0,l}_{\rm{FB}}$, $\mathcal{A}_l$, $\mathcal{A}_c$, $\mathcal{A}_b$, $A^{0,c}_{\rm{FB}}$, $\AbFB$, $R^0_c$, $R^0_b$, $M_W$ and $\Gamma_W$.

The coupling of the $Z$ to left and right handed bottoms, denoted by $g_{Lb}$ and $g_{Rb}$, are given in the following interaction term,
\begin{equation}
\mathcal{L} \supset \frac{g}{c_W} Z_\mu ( g_{Lb} \bar{b}_L \gamma^\mu b_L  + g_{Rb} \bar{b}_R \gamma^\mu b_R)  \,, \label{eq:zbb}
\end{equation}
where $c_W \equiv \cos{\theta_W}$, with $\theta_W$ the weak mixing angle, and $g$ is the $SU(2)$ gauge coupling.  Through out this paper we shall use $\dgLb$ and $\dgRb$ to parameterize the modification of the $Zb\bar{b}$ couplings, defined as
\begin{equation}
g_{Lb} = g^{\rm SM}_{Lb} + \dgLb \, , ~~~ g_{Rb} = g^{\rm SM}_{Rb} + \dgRb \,,  \label{eq:dgb}
\end{equation}
where $g^{\rm SM}_{Lb}$ and $g^{\rm SM}_{Rb}$ are the SM predictions for $g_{Lb}$ and $g_{Rb}$, which at the tree level are given by
\begin{equation}
g^{\rm SM,0}_{Lb} = -1/2 + s^2_W/3 \simeq -0.42 \,, ~~  g^{\rm SM,0}_{Rb} = s^2_W/3 \simeq 0.077  \,.
\end{equation}
At the tree level, $R^0_b$, $\mathcal{A}_b$ and $\AbFB$ at $Z$-pole can be written as
\begin{equation}
R^0_b = \frac{g^2_{Lb} +g^2_{Rb} }{ \underset{q}{\sum}{(g^2_{Lq} +g^2_{Rq})} } \, ,  \label{eq:rbtree}
\end{equation}
where $\underset{q}{\sum}$ denotes a sum over all quarks except the top quark, and
\begin{equation}
\mathcal{A}_b= \frac{ g^2_{Lb} - g^2_{Rb}  }{ g^2_{Lb} +g^2_{Rb}  } \,, ~~~ 
\AbFB = \frac{3}{4}\mathcal{A}_e \mathcal{A}_b  = \frac{3}{4}  \frac{ g^2_{Le} - g^2_{Re}  }{ g^2_{Le} +g^2_{Re}} \frac{ g^2_{Lb} - g^2_{Rb}  }{ g^2_{Lb} +g^2_{Rb}  } \,.   \label{eq:abtree}
\end{equation}
These expressions will be modified once loop corrections are included.

It should be pointed out that the measurements at $Z$-pole alone could not determine the signs of $g_{Lb}$ and $g_{Rb}$.  The off-peak measurements can resolve the sign ambiguities due to interference of the $Z$ diagram with the $s$-channel photon diagram. However, as pointed out in Ref.~\cite{Choudhury:2001hs}, the LEP data at scales different from $m_Z$ are limited in statistics and could definitely resolve the sign of $g_{Lb}$, but not the one of $g_{Rb}$.  Nevertheless, NP theories able to flip the sign of $g_{Rb}$ are typically in tension with other EW precision data such as the constraints on $S$ and $T$ parameters~\cite{Choudhury:2001hs,Batell:2012ca}.  Future lepton colliders will collect a large amount of data at higher scales, which should completely resolve this ambiguity\footnote{As an example, the value of $\AbFB$ at around $240$~GeV is changed by $\sim0.2$ if the sign of $g_{Rb}$ is flipped with respect to the SM prediction. On the other hand, the proposed CEPC run at $\sim 240$~GeV would collect $5~\rm{ab}^{-1}$ of data over ten years with two detectors \cite{CEPCPreCDR}, resulting in a statistical uncertainty of $\sim 0.0003$ for $\AbFB$, which is sufficient for resolving the sign of $g_{Rb}$ as long as enough events are left after the event selection and the systematics are under control.}.  In this paper we do not consider the possibility that $g_{Rb}$ has the opposite sign of its SM prediction.

The potential NP that modifies the $Zb\bar{b}$ couplings can also change other EW observables.  To capture the most relevant corrections without relying too much on the model assumptions, we will consider NP scenarios that contribute to the oblique parameters $S$ and $T$ along with the modified $Zb\bar{b}$ couplings.  Therefore, our minimal model assumption is SM together with $S$, $T$, $\dgLb$ and $\dgRb$ treated as free parameters.  For later convenience we will denote it as (SM$+ S,T,\dgLb, \dgRb$).

\begin{figure}[t]
\centering
\includegraphics[width=7.8cm]{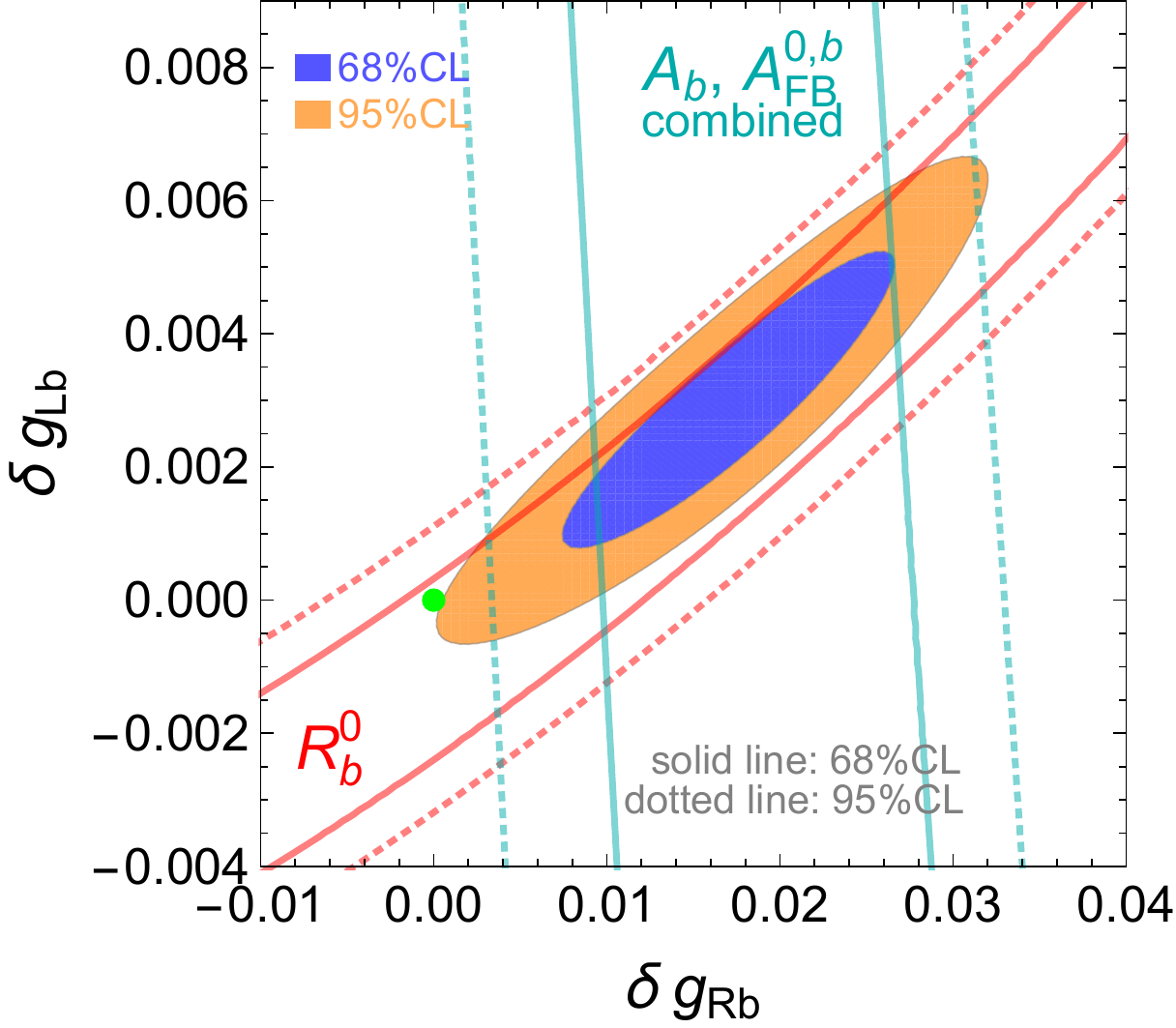} \hspace{0.3cm}
\includegraphics[width=7.5cm]{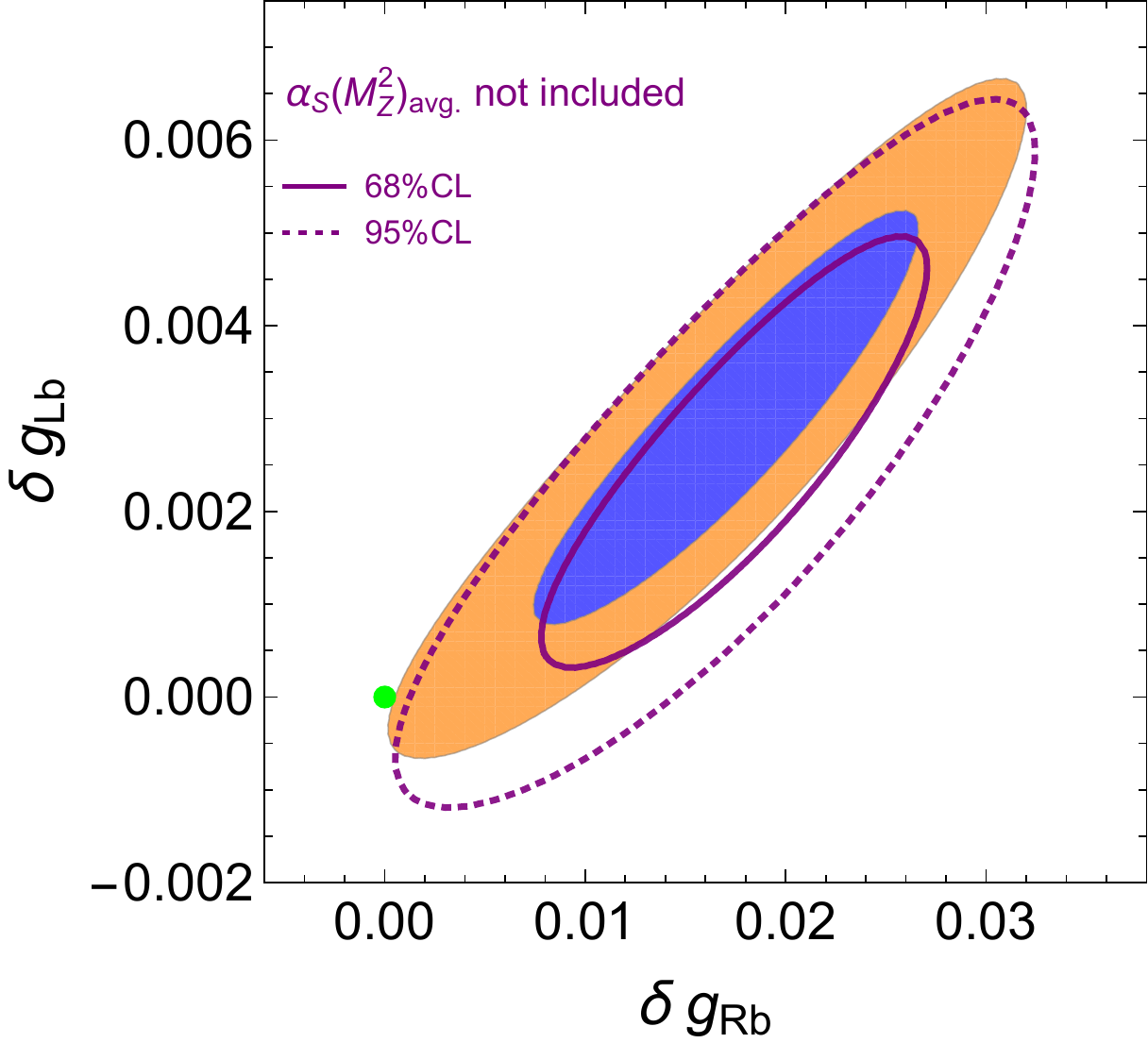} 
\caption{The preferred region in the $(\dgLb, \dgRb)$ plane obtained by the global fit to (SM$+ S,T,\dgLb,\dgRb$) with current data.  The blue (orange) region corresponds to the 68\% (95\%) CL and the green dot is the SM prediction ($\dgLb=\dgRb=0$).  {\bf Left:} The individual constraints from $R^0_b$ and from the combination of $\mathcal{A}_b$ and $\AbFB$ are shown in red and cyan, respectively. For these curves, the parameters other than $\dgLb$ and $\dgRb$ are set to the best-fit values.  {\bf Right:} The purple contours show the preferred regions for which $\asexp$ is not included in the fit.  The solid line corresponds to a 68\% CL and the dotted line corresponds to a 95\% CL. }
\label{fig:z1}
\end{figure}

With the model assumptions and fit procedure described above, we obtain the constraints on $\dgLb$ and $\dgRb$, shown in Fig.~\ref{fig:z1}.  The blue (orange) region corresponds to a confidence level (CL) smaller than 68\% (95\%), while the green dot is the SM prediction ($\dgLb=\dgRb=0$).  In addition, in the left plot we show the individual constraints from $R^0_b$ (red) and the combination of $\mathcal{A}_b$ and $\AbFB$ (cyan), for which the parameters other than $\dgLb$ and $\dgRb$ are set to the best-fit values.  This verifies that  $R^0_b$ and $\AbFB$($\mathcal{A}_b$) are the most relevant measurements for constraining the $Zb\bar{b}$ couplings.  Given that $\dgLb$ and $\dgRb$ are relatively small, the change in $R^0_b$, $\AbFB$ and $\mathcal{A}_b$ can be expanded in terms of $\dgLb$ and $\dgRb$.  Keeping the first order, while fixing the other parameters to the best-fit values, we obtain 
\begin{align}
\delta R^0_b \approx & ~ -0.78\, \dgLb + 0.14\, \dgRb  \, ,  \nonumber \\
\delta \AbFB \approx & ~ -0.034\, \dgLb -0.18\, \dgRb  \approx 0.11 \times \delta \mathcal{A}_b\,.   \label{eq:dRA} 
\end{align}
As shown in Table~\ref{tab:ra0}, the current experimental ($1\sigma$) uncertainties are  $0.00066$ for $R^0_b$ and $0.0016$ ($0.020$) for $\AbFB$ ($\mathcal{A}_b$).  These numbers together with Eq.~(\ref{eq:dRA}) provide a good analytical understanding of the $Zb\bar{b}$ coupling constraints.  First, $R^0_b$ is more constraining than $\AbFB$($\mathcal{A}_b$) and leads to a large positive correlation between $\dgLb$ and $\dgRb$.  Second, $R^0_b$ is numerically more sensitive to $\dgLb$ while $\AbFB$($\mathcal{A}_b$) is more sensitive to $\dgRb$.

As mentioned previously, in our fit we have included $\asexp$ (world average) as a constraint in the global fit.  $\asexp$ includes several different measurements, but is dominated by the lattice calculation \cite{Pich:2013sqa, Agashe:2014kda}.  More explicitly, to avoid double counting, we use the PDG world average excluding electroweak precision test (EWPT) results\cite{Agashe:2014kda}, which is 
\begin{equation}
\asexp = 0.1185 \pm 0.0005 ~~~\mbox{(world average w/o EWPT result)}~.
\end{equation}
The Gfitter group\cite{Flacher:2008zq, Baak:2014ora} does not include this constraint in the fit, since the global fit for the SM and also for the oblique parameters $S$ and $T$ is not sensitive to $\asexp$. One can simply extract the value of $\alpha_S(M^2_Z)$ from the EW global fit, which is in good agreement with $\asexp$ assuming the SM (+$S$ and $T$). This can be seen by our first two results in Table~\ref{tab:alphaS1}.  However, the extraction of $\alpha_S(M^2_Z)$ from EW global fit has some model dependence and, with our model assumption, (SM$+ S,T,\dgLb,\dgRb$), the agreement with $\asexp$ becomes a bit worse (but still $<1\sigma$).  This result is shown in the last row of Table~\ref{tab:alphaS1}.
\begin{table}[t]
\centering
\begin{tabular}{c|c} \hline
model & $\alpha_S(M^2_Z)$ from EW global fit \\ \hline
SM &  $0.1185\pm0.0026$  \\
SM$+(S,T)$ & $0.1180\pm0.0027$ \\
SM$+(S,T,\dgLb,\dgRb)$ & $0.1153\pm0.0035$ \\ \hline
\end{tabular}
\caption{Prediction for $\alpha_S(M^2_Z)$ from EW global fit with different model assumptions and without the world average $\asexp= 0.1185 \pm 0.0005$ as a constraint.}
\label{tab:alphaS1}
\end{table}
This suggests that including $\asexp$ in the fit can have some impact on the $Zb\bar{b}$ coupling constraints.  Indeed, as shown in the right plot of Fig.~\ref{fig:z1}, the inclusion of $\asexp$ has some small but non-negligible effect on $\dgLb$ and on its correlation with $\dgRb$. This is because $\asexp$ has a stronger effect on $R^0_b$ than on $\AbFB$ and prefers smaller $\delta R^0_b$s, leading to an increase in $\delta g_{Lb}$ and a smaller decrease in $\dgRb$ (see Eq.~(\ref{eq:dRA})).  In the future, while both the precision of the $Z$-pole data and $\asexp$ will be improved, $\asexp$ will at least provide an important consistency check and it will be interesting to include it in the global fit for the $Zb\bar{b}$ coupling constraints.

To summarize this Section, in Table~\ref{tab:fitSM0} we list both the individual values of and the correlations among $S,T,\dgLb,\dgRb$ obtained from the global fit of the current precision electroweak data, with model assumption being (SM$+S,T,\dgLb,\dgRb$).  A given NP model may contribute to both $S$, $T$ and $\dgLb$, $\dgRb$, and to constrain the model one should in principle include all four parameters.  However, as shown in Table~\ref{tab:fitSM0}, the correlation between the two groups, $(S,T)$ and $(\dgLb,\dgRb)$, are not very strong, and we expect a similar behavior at future colliders, given that the relative improvements are not extremely different for different observables.  For simplicity, in the next Section we shall focus on the constraint on $\dgLb$ and $\dgRb$ and marginalize over $S$ and $T$.  We refer the reader to other literature, e.g. Ref.~\cite{Fan:2014vta}, for prospective constraints on $S$ and $T$ at future $e^+e^-$ colliders.

%%%%%%%%%%%%%%%%%
\begin{table}
\centering
\begin{tabular}{c|cccc} \hline
 & $S$ & $T$ & $\dgLb$ & $\dgRb$  \\ \hline
$S$ & $-0.047\pm 0.097$ &   &   &   \\
$T$ &  0.91  &  $0.015\pm 0.077$  &  &  \\
$\dgLb$ &  -0.34  &   -0.23  &  $0.0030\pm 0.0015$ &  \\
$\dgRb$ & -0.40 & -0.30 &  0.91 &   $0.0176\pm 0.0063$ \\ \hline
\end{tabular}
\caption{Best fit values ($\pm1\sigma$) of and correlations among $S,T,\dgLb,\dgRb$ from current precision electroweak data, with model assumption being (SM$+S,T,\dgLb,\dgRb$).  A Gaussian distribution is assumed.
}
\label{tab:fitSM0}
\end{table}

%%%%%%%%%%%%%%%%%%%%%%%%%%%%%%%%%%%%%%%%%%%%%%%%%%%%%%%%%%%%%%%%%%%%%%%%%%%%%%
%%%%%%%%%%%%%%%%%%%%%%%%%%%%%%%%%%%%%%%%%%%%%%%%%%%%%%%%%%%%%%%%%%%%%%%%%%%%%%

\section{\boldmath $Zb\bar{b}$ coupling constraints from future $e^+e^-$ colliders}
\label{sec:future}

Future $e^+e^-$ colliders will be able to significantly improve the precision of the measurements at the $Z$-pole thanks to a much larger statistics.   
The reaches have been estimated in the Technical Design Report (TDR) for ILC \cite{Baer:2013cma}, the TLEP whitepaper for FCC-ee\cite{Gomez-Ceballos:2013zzn} and the preliminary Conceptual Design Report (preCDR) for CEPC \cite{CEPCPreCDR}.  However, these estimations usually either contain only a subset of EW observables, or have combined several observables into one ({\it e.g.} the effective leptonic mixing angle $\sin{\theta^l_{\rm eff}}$), and are therefore not straight forward to apply in our study.  In addition, some of the estimations are rather preliminary, having strong dependence on the assumptions for systematic uncertainties and whether or not beam polarization will be implemented.   
 In Section~\ref{sec:futin}, we outline the key observables that are needed for improving the $Zb\bar{b}$ coupling constraints and try to estimate their precision reach at the three future colliders.   
Using these estimations, we proceed to study the constraints on the $Zb\bar{b}$ couplings by the method of global fit, and the results are shown in Section~\ref{sec:fitSM} and \ref{sec:fitA}.

In our study we consider the following benchmark scenarios for the three colliders:
\begin{itemize}
\item  {\bf CEPC} with a relative conservative estimation for the systematic uncertainties and with a statistics of only $2\times10^9$ $Z$ events.  While beam polarization could be a potential option for the run at the $Z$-pole, here we assume that it is not implemented. 
\item  {\bf CEPC+}, which is CEPC with a more aggressive estimation for the systematic uncertainties, and assuming  $10^{10}$ $Z$ events.  
\item  {\bf ILC},  with a lower statistics ($10^9$ $Z$ events), but with beam polarization.
\item  {\bf FCC-ee} with $10^{12}$ $Z$ events and beam polarization.
\end{itemize}

\subsection{Precision of the EWPOs at future \boldmath $e^+e^-$ colliders}
\label{sec:futin}

The observables directly related to the $Zb\bar{b}$ couplings are $R^0_b$, $\AbFB$ (measured without beam polarization) and $\Ab$ (measured with beam polarization).   However, the three observables also have explicit dependence on the effective weak mixing angles, $\sin^2{\theta^l_{\rm eff}}$ (for leptons) and $\sin^2{\theta^b_{\rm eff}}$ (for bottom)\footnote{This can be seen from the tree-level expressions in Eq.~(\ref{eq:rbtree}) and Eq.~(\ref{eq:abtree}).  The effective weak mixing angle for leptons ($\sin^2{\theta^l_{\rm eff}}$) and bottom ($\sin^2{\theta^b_{\rm eff}}$) are not exactly the same due to different loop contributions.}. In particular,  $\AbFB$ is quite sensitive to $\sin^2{\theta^l_{\rm eff}}$ as it is proportional to $\Ae$.  In fact, at present, the LEP measurement of $\AbFB$ provides one of the best determination of $\sin^2{\theta^l_{\rm eff}}$ assuming SM (the other one being $A_{LR}$ from SLD).  Therefore, to extract the $Zb\bar{b}$ couplings from $\AbFB$, it is important to obtain an independent determination of $\sin^2{\theta^l_{\rm eff}}$, while the most precise such determination is provided by the leptonic asymmetry observables.  On the other hand, $R^0_b$ and $\Ab$ are numerically not very sensitive to $\sin^2{\theta^b_{\rm eff}}$.  

Without beam polarization, the forward-backward leptonic asymmetry $A^{0,l}_{\rm FB} \,(=\frac{3}{4}\Al^2)$ can be measured.  In addition, a measurement of $\Al$ ($\Ae$ and $\mathcal{A}_\tau$) can be obtained using the average final-state longitudinal $\tau$ polarization and its forward-backward asymmetry\cite{ALEPH:2005ab}, which we denote as $\Al (\mathcal{P}^{\rm}_\tau)$.  With beam polarization, the left-right asymmetry $A_{LR} \,(= \Ae)$ can be directly measured, but it is rather irrelevant in terms of the $Zb\bar{b}$ coupling constraints as $\Ab$ can be directly measured as well, and we have checked that the impact of the improvement of $A_{LR}$ on the $Zb\bar{b}$ coupling is rather negligible.  However, $A_{LR}$ can still be helpful for constraining the $Zb\bar{b}$ couplings in a global fit, for example, in the case that two (or more) colliders are built and only one of them have beam polarization, similar to the situation of LEP and SLC. 

It is worth noting that, apart from $R^0_b$, a number of additional observables are also sensitive to the coupling combination $g^2_{Lb}+g^2_{Rb}$ through the dependence on the total hadronic decay width, among which $R^0_l$, which is the ratio of the total hadronic $Z$ decay width and the $Z$ decay width to one lepton species, is relatively well measured and provides the best sensitivity.    
We find that a significant improvement of the precision of $R^0_l$ can have a significant impact on the $Zb\bar{b}$ coupling constraints. In particular, with the estimated precision reach at FCC-ee shown later, the measurement of $R^0_l$ turns out to be more constraining than the one of $R^0_b$ to the $Zb\bar{b}$ couplings.  However, the constraint from $R^0_l$ depends strongly on the assumption that 
the coupling of $Z$ to other fermions are SM-like, and one should be cautious when applying the $Zb\bar{b}$ coupling constraints to a model for which this assumption is not true. In the end of Section \ref{sec:fitSM} we will also show the results for FCC-ee with a more conservative estimation for the precision of
$R^0_l$.

To obtain the estimation of the precision reach of the several observables, the following procedure is performed.   For each observable, we use the estimation in the corresponding literature, if it is provided. In particular, if a range of values is provided, we choose the more conservative one.  If the estimation is not provided in the literature, we estimate the precision with the following strategies: we assume that the systematic uncertainties at CEPC is a factor of $1/3$ the ones at LEP and the systematic uncertainties for the scenario ``CEPC+" is reduced by an additional factor of $1/2$  from  CEPC\footnote{An exception of this is the CEPC systematic uncertainty for Rl0(0.006), which we deduced from the total uncertainty (0.007) in the preCDR \cite{CEPCPreCDR} and then scale to obtain the systematic uncertainty at CEPC+ (0.003).}.  For CEPC and CEPC+, we assume the statistical uncertainty simply scales with $1/\sqrt{N}$, where $N$ is the total number of $Z$ events expected to be collected. 
Additionally, for ILC, Ref.~\cite{Baer:2013cma} does not provide an estimation for the uncertainty of $R^0_l$ ($\Delta R^0_l$),   
for which we adopt the estimation in Ref.~\cite{Flacher:2008zq} by Gfitter, $\Delta R^0_l=0.004$.   
Finally, for the FCC-ee, Ref.~\cite{Gomez-Ceballos:2013zzn} does not provide an estimation for the uncertainty of $\Ab$ ($\Delta \Ab$).  We na\"ively scale it from the estimation for the ILC, assuming $\frac{ \Delta \Ab }{ \Delta A_{LR} }|_{\rm ILC} \approx \frac{ \Delta \Ab }{ \Delta A_{LR} }|_{\rm FCC-ee}$, which gives $\Delta \Ab \approx 0.00021$ at FCC-ee.  

\begin{table}[t]\small
\centering
\begin{tabular}{|c||c|c|c|c|c|} \hline
& \multicolumn{5}{c|}{Precision}  \\ \hline
 Observable & Current &  CEPC & CEPC+ & ILC & FCC-ee   \\ \hline\hline
$R^0_b$ & 0.00066   & 0.00017  & \hl{0.00008} & 0.00014 & 0.00006    \\
$ [0.21629]$ & (0.00050) & (0.00016) & \hl{(0.00008)} &  & (0.00006)    \\ \hline 
$R^0_l$ & 0.025  & 0.007  & \hl{0.003} & 0.004 & 0.001 \\
$[20.767]$& (0.007) & \hl{(0.006)} &  \hl{(0.003)} &    & (0.001)  \\\hline\hline  
 $\AbFB$ & 0.0016  & 0.00015  & \hl{0.00007} &  &   \\
$[0.0992]$&  (0.0007)  &  (0.00014)  & \hl{(0.00007)}  &  &    \\ \hline 
 $\AlFB$  &  0.0010 & \hl{0.00014 } & \hl{0.00007} &  &  \\
$[0.0171]$ &   (0.0003)   &  \hl{(0.00010)}  & \hl{(0.00005)} &   &      \\ \hline
 $\Al (\mathcal{P}^{\rm}_\tau)$  & 0.0033 & \hl{0.0006} &  \hl{0.0003} &  &   \\
$[0.1465]$ & (0.0015) & \hl{(0.0005)} & \hl{(0.0003)}  &    &    \\ \hline\hline
  $\Ab$ & 0.020 &  &   & 0.001 &  \hl{0.00021}    \\
 $[0.923]$ &    &    &    &    & \hl{(0.00015)}  \\ \hline 
 $A_{LR}$ & 0.0022 &  & & 0.0001 & 0.000021    \\
$[0.1514]$   & (0.0011) &    &    &  (0.0001)  & (0.000015)   \\ \hline \hline
 \# of $Z$s & $\sim2\times10^7$ & $\sim2\times10^9$ & $\sim10^{10}$ & $\sim10^9$ & $\sim10^{12}$  \\\hline
\end{tabular}
\caption{
The estimated precision reach  for the observables (which current experimental measurement is shown in the first column) most relevant to constrain the $Zb\bar b$ coupling at future colliders.  The second column shows the uncertainty of the present measurements from LEP and SLC, while the other columns show the estimations of the precision reach for different future colliders and scenarios.  The numbers highlighted in \hl{blue} are our own estimation.  In each entry, the number at the top shows the total uncertainty while the number at the bottom (in parenthesis) shows the corresponding systematic uncertainty.  A blank entry denotes an observable that is either not measured or not important for our global fit.  The last row shows the expected number of $Z$ events that will be collected. 
}
\label{tab:inputx}
\end{table}

The estimations for the observables mentioned above are summarized in Table~\ref{tab:inputx}.  A similar method is used to estimate the precision reach for the additional EW observables not listed in Table~\ref{tab:inputx}, even if we have checked that they have a much smaller impact on the $Zb\bar{b}$ coupling constraints.
In Table~\ref{tab:inputx}, the numbers highlighted in \hl{blue} are our own new estimations for those observables not reported in the literature.  In each entry, the number at the top shows the total uncertainty while the number at the bottom (in parenthesis) shows the corresponding estimated systematic uncertainty\footnote{For some of the observables, only the total experimental uncertainties are quoted in the literature and the entries for the systematic uncertainties are left blank, accordingly.}.  For comparison, in the second column of the table, we show the current uncertainties, taken from Ref.~\cite{ALEPH:2005ab}. A few comments on the table are in order:  the systematic uncertainties mostly dominate except for $\Ab$ and $A_{LR}$ at FCC-ee, and for $\AlFB$ at CEPC.  Without beam polarization, CEPC can not measure $\Ab$ and $A_{LR}$. For this reason, the corresponding entries are left blank. ILC and FCC-ee can measure $\AbFB$, $\AlFB$ and  $\Al (\mathcal{P}^{\rm}_\tau)$, but the corresponding observable with beam polarization, $\Ab$ and $A_{LR}$, can be measured significantly more precisely, so, for simplicity, we do not include the former observables.

A potential issue for the interpretation of future measurements is the effects of the theoretical uncertainties, which could become important if they are much larger than the experimental uncertainties.  In our study, we assume that the electroweak three-loop corrections will be computed in the future.  In that case, the effects from the theoretical uncertainties on the $Zb\bar{b}$ coupling constraints are very small and can be safely neglected even for FCC-ee.  This is either because the theoretical uncertainty is numerically small (such as $\delta_{\rm th} R^0_b$ which is estimated to be a few times $10^{-5}$\cite{Fan:2014axa}) or the observable itself has little effect on the $Zb\bar{b}$ coupling constraints, such as the top quark mass.  The theoretical uncertainty of $\sin^2{\theta^b_{\rm eff}}$ also has little impact, since $\Ab$ is not very sensitive to it. More details on the treatment of the theoretical uncertainties can be found in Appendix~\ref{app:theo}.

%%%%%%%%%%%%%%%%%%%%%%%%%%%%%%%%%%%%%%%%%%%%%%%%%%%%%%%%
%%%%%%%%%%%%%%%%%%%%%%%%%%%%%%%%%%%%%%%%%%%%%%%%%%%%%%%%

\subsection{SM-like measurements and constraints on NP}
\label{sec:fitSM}

\begin{figure}[t]
\centering
\includegraphics[width=7.3cm]{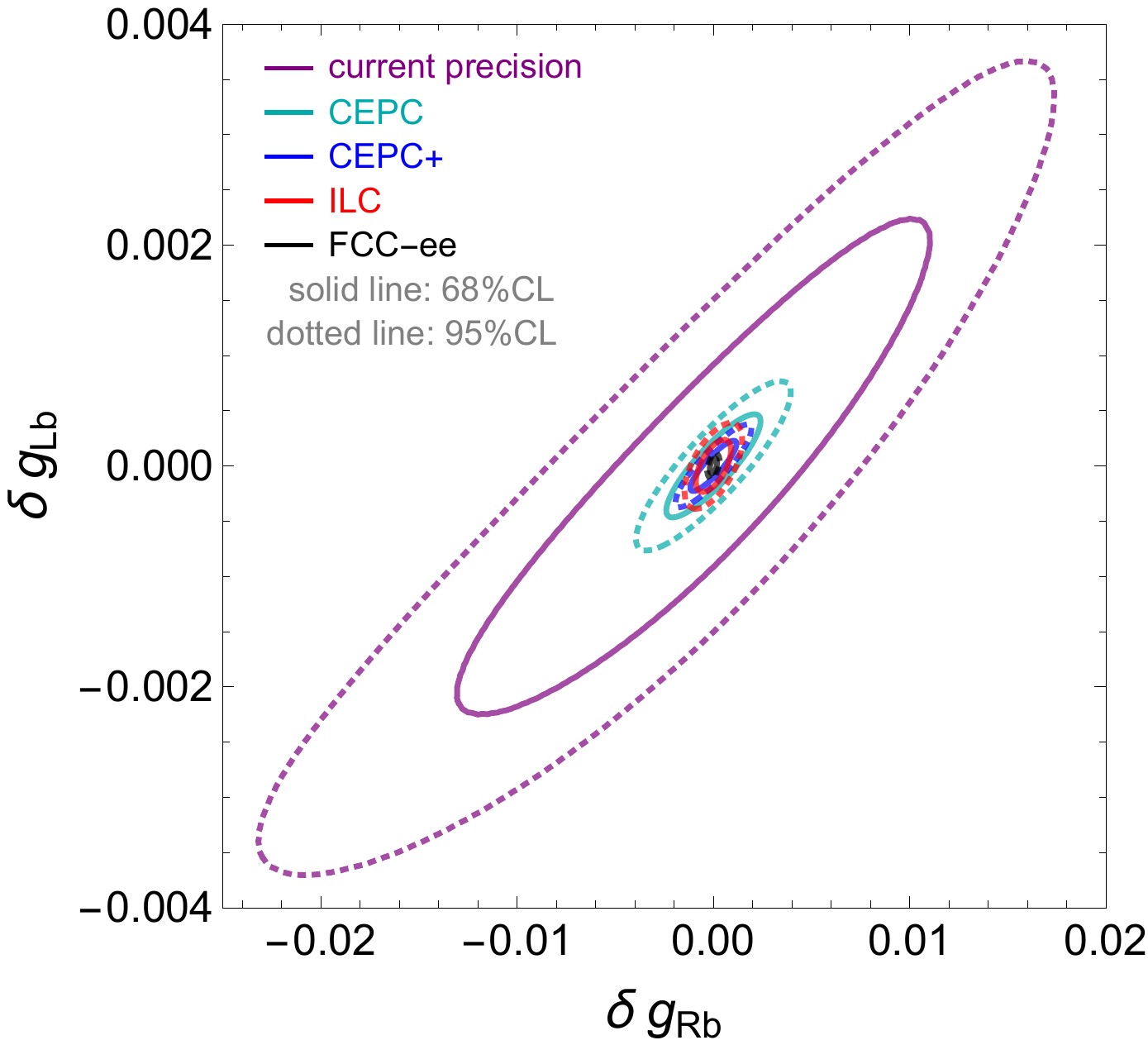} \hspace{0.3cm}
\includegraphics[width=7.6cm]{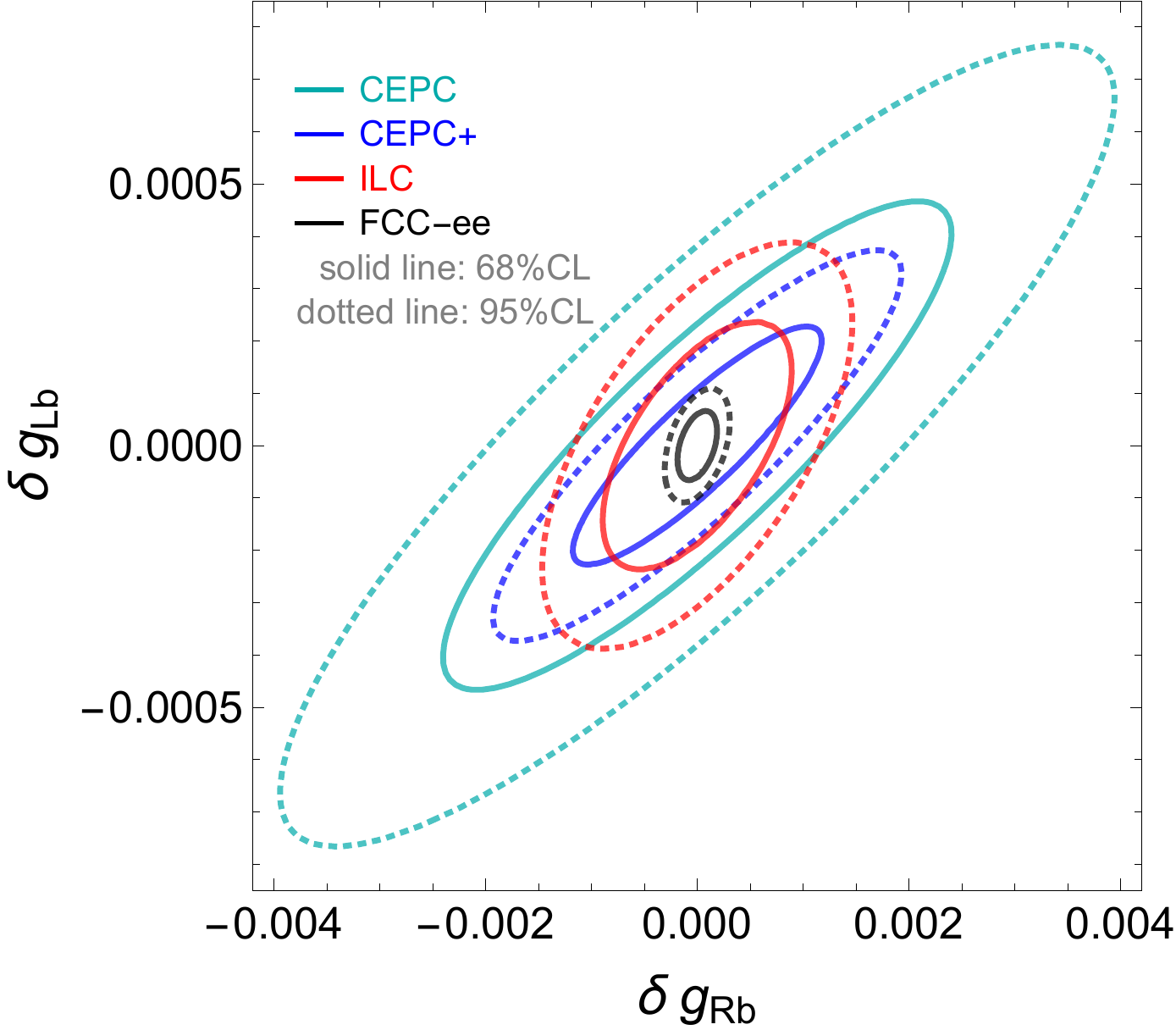}
\caption{Preferred regions in the $(\dgLb,\dgRb)$ plane, assuming SM central values for all measurements.  The model assumption is (SM$+S,T,\dgLb,\dgRb$), with $S,T,\dgLb,\dgRb$ all treated as free input parameters.  The solid and dotted lines are 68\% and 95\% CLs, respectively.  The purple contours assume current precision for all measurements.  The cyan, blue, red and black contours correspond to the estimated precisions for CEPC, CEPC+, ILC and FCC-ee, respectively.  The plot in the right panel is a zoomed-in version of the one in the left panel.
}
\label{fig:ccif}
\end{figure}

In this Section, we assume the future experimental results agree perfectly with the SM predictions and the estimated precision of future measurements as described in the previous Section.  The preferred regions in the $(\dgLb, \dgRb)$ plane obtained by our global fit are shown in Fig.~\ref{fig:ccif}.  The plot in the right panel is a zoomed-in version of the one in the left panel.

From Fig.~\ref{fig:ccif}, it is clear that the constraints on the $Zb\bar{b}$ coupling are significantly improved at the future $e^+e^-$ colliders, even for the relatively conservative CEPC estimation (cyan contours), compared to the results of the current precisions (purple contours).  With beam polarization, ILC (red contours in the figure) and FCC-ee (black contours in the figure) have better measurements of $\Ab$, which gives a better constraint on $\dgRb$ and also reduce the correlation between $\dgLb$ and $\dgRb$.

We report the $1\sigma$ uncertainties of $\dgLb$ and $\dgRb$ as well as their correlation ($\rho$) in Table~\ref{tab:ccif}.  Due to the strong correlation between $\dgLb$ and $\dgRb$ (in particular at CEPC), one need to be careful when using these results to constrain NP models, since in some models only one of $\dgLb$ and $\dgRb$ can receive a sizable contribution while the other one is close to zero.  Therefore, in Table~\ref{tab:ccif} we also show the $1\sigma$ uncertainties for $\dgLb$($\dgRb$), while $\dgRb$($\dgLb$) is fixed to zero.

\begin{table}[t]
\centering
\begin{tabular}{|c||c|c|c||c|c|} \hline
& $\dgLb$ & $\dgRb$ & $\rho$ & $\dgLb~(\dgRb=0)$ & $\dgRb~(\dgLb=0)$  \\ \hline\hline
current & 0.0015 & 0.0079 & 0.91  & 0.00061  & 0.0032  \\ \hline 
CEPC & 0.00031 & 0.0016 &  0.87 & 0.00015  & 0.00079  \\ \hline 
CEPC+ & 0.00015  & 0.00078  & 0.88  &  0.000072 &  0.00037  \\ \hline 
ILC & 0.00016  & 0.00059  & 0.61  & 0.00012 & 0.00047  \\ \hline 
FCC-ee & 0.000044  & 0.00012 & 0.42  &  0.000040 & 0.00011   \\ \hline 
\end{tabular}
\caption{A comparison of precision reach at different future colliders.  The 2nd(3rd) column shows the $1\sigma$ uncertainties of $\dgLb$($\dgRb$) while marginalizing over $\dgRb$($\dgLb$).  The 4th column shows the correlation ($\rho$) between $\dgLb$ and $\dgRb$.  The 5th(6th) column shows the $1\sigma$ uncertainty of $\dgLb$($\dgRb$) with $\dgRb$($\dgLb$) fixed at zero.  We assume future measurements to be in perfect agreement with the SM predictions. A Gaussian distribution is assumed.}
\label{tab:ccif}
\end{table}

For CEPC, the estimation for $\Delta R_l^0$ in the preCDR \cite{CEPCPreCDR} seems to be very conservative, suggesting little improvement of its systematic uncertainty from LEP to CEPC (see Table \ref{tab:inputx}). A scaling of a factor of 1/3 on the systematic uncertainty would give a value of $\Delta R_l^0 = 0.003$ for the total uncertainty. The results for this scenario are shown in Table \ref{tab:ccif003}, which exhibits a slight improvement. In Table \ref{tab:ccif003} we also show the results for FCC-ee with a more conservative estimation of $\Delta R_l^0$, also using $\Delta R_l^0 = 0.003$ (instead of $\Delta R_l^0 = 0.001$).

\begin{table}[t]
\centering
\begin{tabular}{|c||c|c|c||c|c|} \hline
$\Delta R_l^0 = 0.003$ & $\dgLb$ & $\dgRb$ & $\rho$ & $\dgLb~(\dgRb=0)$ & $\dgRb~(\dgLb=0)$  \\ \hline\hline
CEPC & 0.00031 & 0.0015 & 0.93  & 0.00011  & 0.00057  \\ \hline 
FCC-ee & 0.000051  & 0.00012 & 0.32  &  0.000048 & 0.00012   \\ \hline 
\end{tabular}
\caption{Same as Table \ref{tab:ccif}, but for CEPC and FCC-ee both with $\Delta R_l^0 = 0.003$, which serves as a reasonably optimistic estimation for CEPC and a conservative one for FCC-ee.}
\label{tab:ccif003}
\end{table}

%%%%%%%%%%%%%%%%%%%%%%%%%%%%%%%%%%%%%%%%%%%%%%%%%%%%%%%%
%%%%%%%%%%%%%%%%%%%%%%%%%%%%%%%%%%%%%%%%%%%%%%%%%%%%%%%%

\subsection{Discovering NP through \boldmath $\AbFB$ ($\Ab$)}
\label{sec:fitA}

A more interesting possibility is that the long standing $\AbFB$ discrepancy does come from NP, in which case the precision reach at any of the three future $e^+e^-$ colliders should be able to rule out the SM with very high significance and therefore provide strong indirect evidence for physics beyond SM.  To illustrate this point, we consider the following two scenarios. {\it{Scenario I}}: we assume that 
the true values for $\dgLb$ and $\dgRb$ (denoted by $\delta g^0_{Lb}$ and $\delta g^0_{Rb}$) are given by the best fit values of the current data, $\delta g^0_{Lb}=0.0030$ and $\delta g^0_{Rb}=0.0176$ (see Table~\ref{tab:fitSM0}). {\it{Scenario II}}: we assume that the the true values of $\delta g^0_{Lb}$ and $\delta g^0_{Rb}$ are closer to zero, while still being consistent with the current measurements within $68\%$CL.  As a benchmark point, we choose $\delta g^0_{Lb}=0.0009$ and $\delta g^0_{Rb}=0.0075$.  In principle, one would expect the NP to have non-zero contributions to the $S$ and $T$ parameters, as well.  We find that changing the central values of $S$ and $T$ of the hypothetical measurement within the current constraints has very small impact on the $Zb\bar{b}$ couplings.  For simplicity we assume that the hypothetical data agrees with SM other than the modification to $g_{Lb}$ and $g_{Rb}$.

\begin{figure}[t]
\centering
\includegraphics[width=7.3cm]{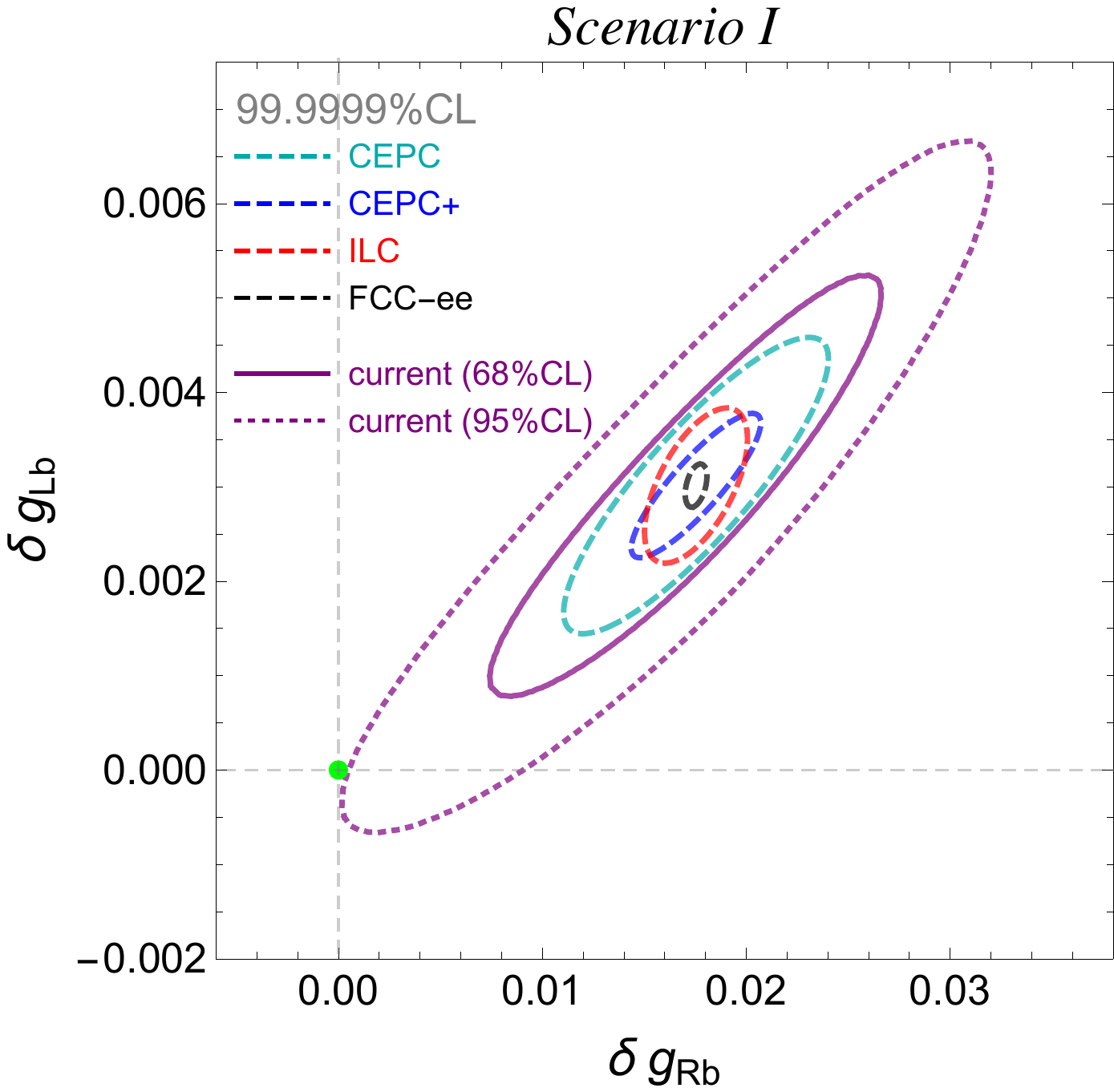} \hspace{0.3cm}
\includegraphics[width=7.3cm]{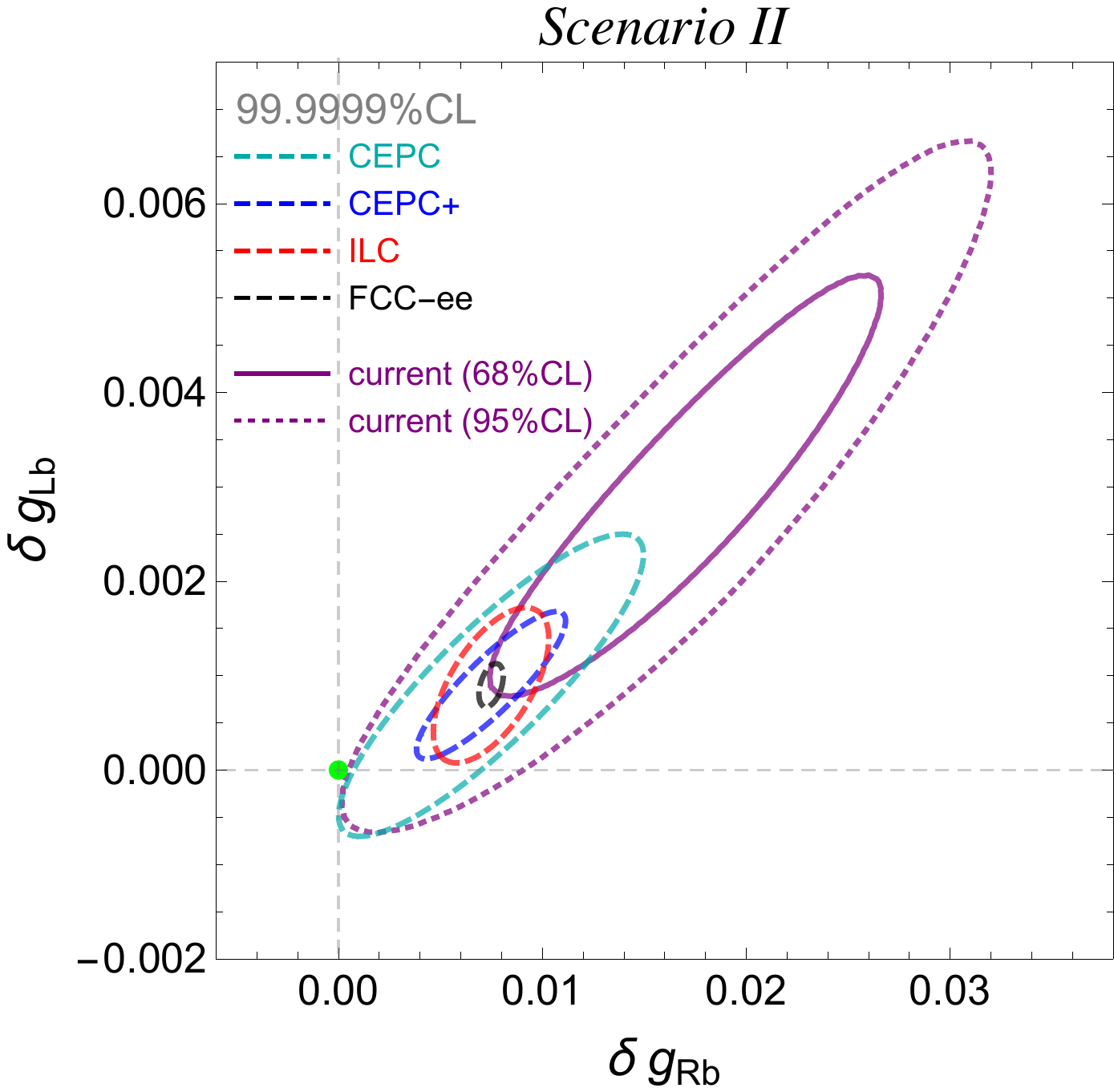}
\caption{ 
The preferred regions in the $(\dgLb,\dgRb)$ plane, given by the global fit of the future measurements at CEPC (in cyan), CEPC+ (in blue), ILC (in red) and FCC-ee (in black).  The solid and dotted purple contours correspond to the 68\% and 95\% CL constraints from the current measurements.  The two panels correspond to {\it{Scenario I}} and {\it{Scenario II}} presented in the text, and each plot shows the $99.9999\%$~CL constraints from different colliders with dashed contours.  The green dot is the SM prediction ($\dgLb=\dgRb=0$).
}
\label{fig:np1}
\end{figure}

The preferred regions in the $(\dgLb,\dgRb)$ plane are shown in Fig.~\ref{fig:np1}.  The two plots correspond to the two scenarios described above, and each shows the $99.9999\%$~CL (corresponding to $5\sigma$ for a one-dimensional Gaussian distribution) constraints from different colliders.  From the figure, it is clear that the SM prediction at zero (denoted by a green dot) can be ruled out at $99.9999\%$~CL by all the $e^+e^-$ colliders we discuss, even if we assume that the future measurements will point towards smaller values of $\dgLb$ and $\dgRb$ within $68\%$~CL of the current measurements.

%%%%%%%%%%%%%%%%%%%%%%%%%%%%%%%%%%%%%%%%%%%%%%%%%%%%%%%%
%%%%%%%%%%%%%%%%%%%%%%%%%%%%%%%%%%%%%%%%%%%%%%%%%%%%%%%%

\section{Implication on NP models}
\label{sec:model}

In this Section, we analyze the implications of the future measurements of the $Zb\bar b$ couplings on specific NP models.  We start with a brief discussion of the constraints on effective Lagrangians. At dimension 6, the only operators that modifies directly the $Zb\bar{b}$ couplings are (see e.g.~\cite{Ciuchini:2013pca,Falkowski:2014tna})
\begin{align}
\mathcal{O}_{Hb} = & ~ i(H^\dagger \overset{\leftrightarrow}{D_\mu} H)(\bar{b}_R \gamma^\mu b_R) \,, \\
\mathcal{O}^s_{HQ} = & ~ i(H^\dagger \overset{\leftrightarrow}{D_\mu} H)(\bar{Q} \gamma^\mu Q)\,,  \\
\mathcal{O}^t_{HQ} = & ~ i(H^\dagger \sigma^a \overset{\leftrightarrow}{D_\mu} H)(\bar{Q} \gamma^\mu \sigma^a Q) \,.
\end{align}
After electroweak symmetry breaking, these operators lead to a shift in the $Zb\bar b$ couplings:
\begin{equation}
\dgLb = -\frac{(a^s_{HQ}+a^t_{HQ})v^2}{2} \,, ~~~~~ \dgRb = -\frac{a_{Hb} v^2}{2} \,,   \label{eq:dgop}
\end{equation}
where $a_{Hb},\,a^s_{HQ},\,a^t_{HQ}$ are the coefficients of the $\mathcal{O}_{Hb},\,\mathcal{O}^s_{HQ},\,\mathcal{O}^t_{HQ}$ operators, respectively and $v$ is the vacuum expectation value of the Higgs ($v=246$ GeV). In Table~\ref{tab:higherDim}, we present the constraints on these operators at the several future $e^+e^-$ machines, assuming that $a^s_{HQ}=a^t_{HQ}=a_{Hb}=1/\Lambda^2$. Scales as large as $(20-30)$~TeV can be probed by the future measurement of the $Zb\bar b$ couplings.

\begin{table}[t]
\centering
\begin{tabular}{c|ccccc} 
& current & CEPC & CEPC+ & ILC & FCC-ee  \\ \hline
$\Lambda$(TeV) &  6.8 & 13 & 20 & 15 & 27 \\
\end{tabular}
\caption{$95\%$CL bounds on the cutoff scale $\Lambda$ from different future colliders, using the results from Section~\ref{sec:fitSM} and assuming that $a^s_{HQ}=a^t_{HQ}=a_{Hb}=1/\Lambda^2$.}
\label{tab:higherDim}
\end{table}

Next, we pass to the analysis of specific NP frameworks that can generate some of the operators forementioned, including two Higgs doublet models, composite Higgs models and the Beautiful Mirror Model.  It should be noted that for natural SUSY with minimal ingredients, the $Zb_L\bar{b}_L$ coupling receives corrections from loops involving stops and Higgsinos, while they are rather small for the future measurements of the $Zb\bar{b}$ to have an impact, as shown e.g. in Ref.~\cite{Fan:2014axa}.

%%%%%%%%%%%%%%%%%%%%%%%%%%%%%%%%%%%%%%%%%%%%%%%%%%%%%%%%

\subsection{Two Higgs doublet models}

\begin{figure}[t]
\centering
\includegraphics[width=0.44\textwidth]{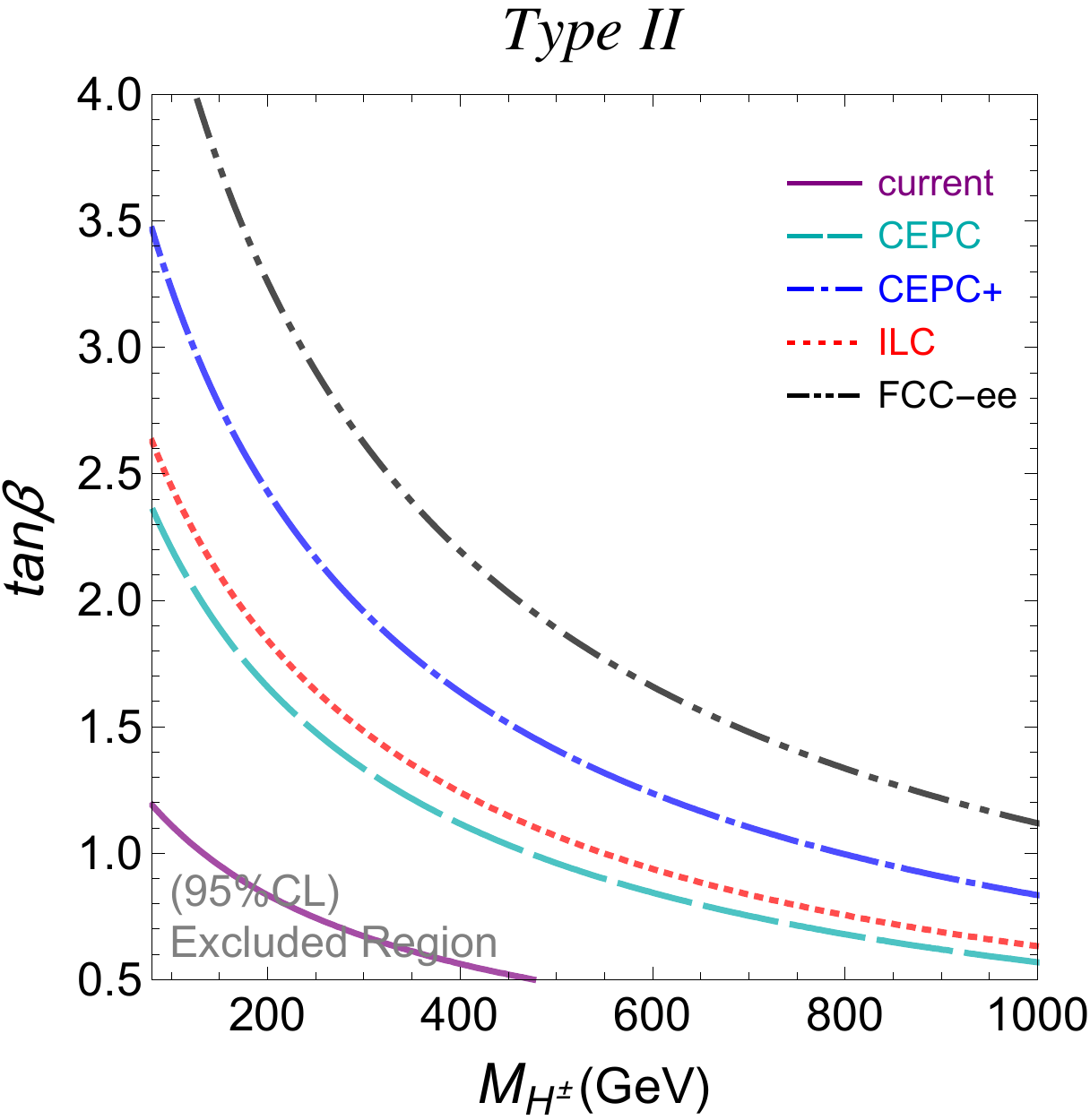} \hspace{0.3cm}
\includegraphics[width=0.44\textwidth]{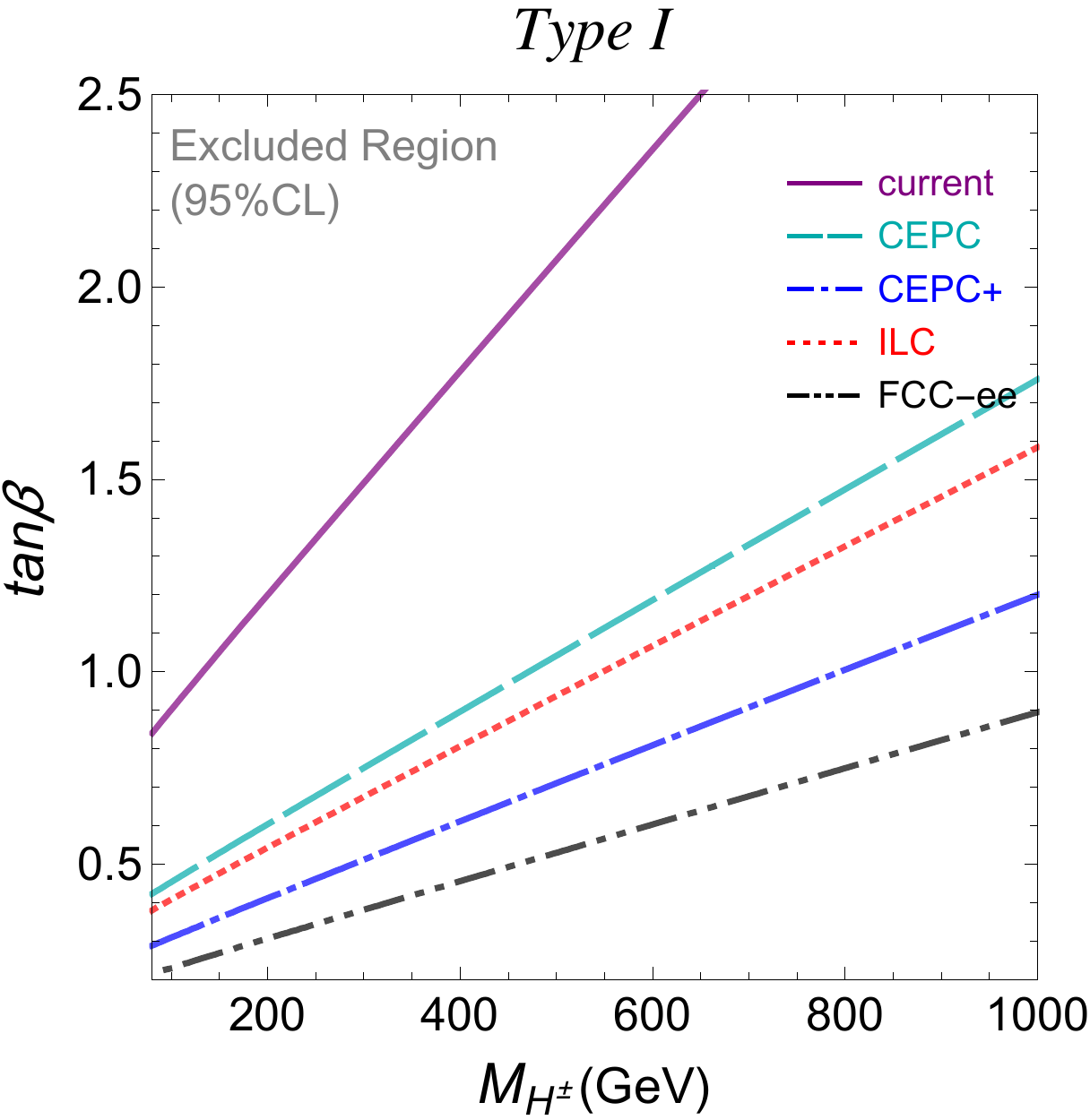}
\caption{Constraints on the charged Higgs parameter space in a Type II 2HDM (left panel) and in a Type I 2HDM (right panel). In purple is the constraint we obtain using the present uncertainties on the EWPOs; in cyan, blue, red and black are the constraints expected with the future measurements at CEPC, CEPC+, ILC and FCC-ee, respectively. In the Type II 2HDM, the region below the curves is excluded. In the Type I 2HDM, the exclusion is above the curves. We assume that the future measurements perfectly agree with the SM predictions and marginalized over the values of the $S$ and $T$ parameters.}
\label{fig:ChargedHiggs}
\end{figure}

As shown in e.g.~\cite{Haber:1999zh}, models with an extended Higgs sector can predict sizable NP contributions to the $Zb\bar b$ vertex. In particular, focusing on two Higgs doublet models (2HDMs), based on discrete symmetries to avoid flavor changing neutral currents (FCNCs) at the tree level, the most important contribution generically comes at the one loop level, from the charged Higgs exchange. The sign of the charged Higgs NP contribution to $\dgLb$ ($\dgRb$) is fixed and is always positive (negative). In a Type II 2HDM, the contribution to $\dgLb$ ($\dgRb$) increases at small (large) values of $\tan\beta$, since the coupling $H^\pm \bar b_L t_R$ ($H^\pm \bar t_L b_R$) leading to a non-zero $\delta g_{Lb}$ ($\delta g_{Rb}$) is proportional to $m_t/\tan\beta$ ($m_b \tan\beta$). In a Type I 2HDM, instead, both $\dgLb$ and $\dgRb$ increase at large values of $\tan\beta$\footnote{Here we use the $\tan\beta$ convention such that the two charged Higgs couplings are proportional to $m_t \tan\beta$ and $m_b \tan\beta$.}, leading always to a NP contribution $\dgLb\gg \dgRb$.
In Fig.~\ref{fig:ChargedHiggs}, we show the constraints on the $m_{H^\pm}-\tan\beta$ plane, using the present measurement of the $Z b \bar b$ coupling (in purple) as well as the expected more precise measurement at CEPC (in cyan), CEPC+ (in blue), ILC (in red) and FCC-ee (in black). For the figure, we have assumed that the future measurements perfectly agree with the SM predictions and we have marginalized over the values of the $S$ and $T$ parameters.

 In Type II models, interesting constraints arise at low values of $\tan\beta$ for which $\delta g_{Lb}\gg \delta g_{Rb}$\footnote{In a Type II 2HDM, only very large values of $\tan\beta$ can be excluded by the measurement of $\delta g_{Rb}$: $\tan\beta\sim \mathcal O(50)$ ($30$) for $m_{H^\pm}\sim 200$ GeV at CEPC+ (FCC-ee). At large $\tan\beta$, the exact exclusion depends also on the neutral Higgs spectrum, since in this case the neutral and charged Higgses give a parametrically similar contribution to $\delta g_{Lb}$ and $\delta g_{Rb}$.}. Type I models, instead, are only allowed in the region with small $\tan\beta$ unless $m_{H^\pm}$ is very large. If we specify the full spectrum of a 2HDM, including the masses of the neutral scalar $H$ and pseudoscalar $A$, as well as the mixing angle $\alpha-\beta$ between the two doublets, $S$ and $T$ are not free parameters. In general, the constraints will be stronger. As an example, in a type II 2HDM, if we fix $m_A=m_H=200$ GeV, $m_{H^{\pm}}=150$ GeV and $\cos(\alpha-\beta)=0.1$, then the model gives $S\sim 0.02$ and $T\sim 0.04$. For this specific set of parameters, the CEPC bound on $\tan\beta$ is  $\sim 2$, instead of $\sim 1.8$, as presented in the left panel of Fig. \ref{fig:ChargedHiggs} (see the cyan dashed line).

Presently, LHC charged Higgs searches almost totally exclude charged Higgs bosons with a mass below the top mass in Type II 2HDMs~\cite{Aad:2014kga}. There are no LHC searches at around the top mass, for $160\,{\rm{GeV}}<m_{H^\pm}<180$ GeV up to date. Above 180 GeV, constraints are rather weak and cover only models with large values of $\tan\beta$ ($\tan\beta\geq 40$), for which the production cross section of the charged Higgs in association with a top quark is in the $\mathcal O(1)$ pb range. In this regime, the two most important bounds come from the searches for $H^\pm\to\tau\nu$~\cite{Aad:2014kga} and for $H^\pm\to t b$~\cite{CMS:2014pea}. At the 14 TeV LHC, also charged Higgs boson with mass above the top mass will be relatively well probed. In particular, searches for $H^\pm\to tb$ will have the potential to probe $\tan\beta\lesssim 3$ and $\tan\beta\gtrsim 15$ for $m_{H^\pm}\sim 500$ GeV at the High Luminosity (HL)-LHC~\cite{Craig:2015jba}.
Comparing to our results of Fig.~\ref{fig:ChargedHiggs} (left panel), we see that constraints from future measurements of EWPOs can be complementary to direct searches for Type II 2HDMs, being able to probe low values of $\tan\beta$ even for $m_{H^\pm}\gg m_t$, as well as the challenging region $160\,{\rm{GeV}}<m_{H^\pm}<180$ GeV, presently not covered by direct searches. 

Finally, one can interpret the searches for charged Higgs bosons in terms of Type I 2HDMs. Below the top mass, only a small region with $\tan\beta<1$ has not been yet probed by the $H^\pm\to\tau\nu$ search. Above the top mass, the exclusion is very week and is not covering any part of the plane shown in Fig.~\ref{fig:ChargedHiggs} (right panel). At the HL-LHC, this region will be very well probed by a $H^\pm \to tb$ search, with potential exclusions for the entire range of $\tan\beta$ presented in the figure, up to $m_{H^\pm}\sim 500$ GeV.

%%%%%%%%%%%%%%%%%%%%%%%%%%%%%%%%%%%%%%%%%%%%%%%%%%%%%%%%

\subsection{Composite Higgs models}

\begin{figure}[t]
\centering
\includegraphics[width=7.3cm]{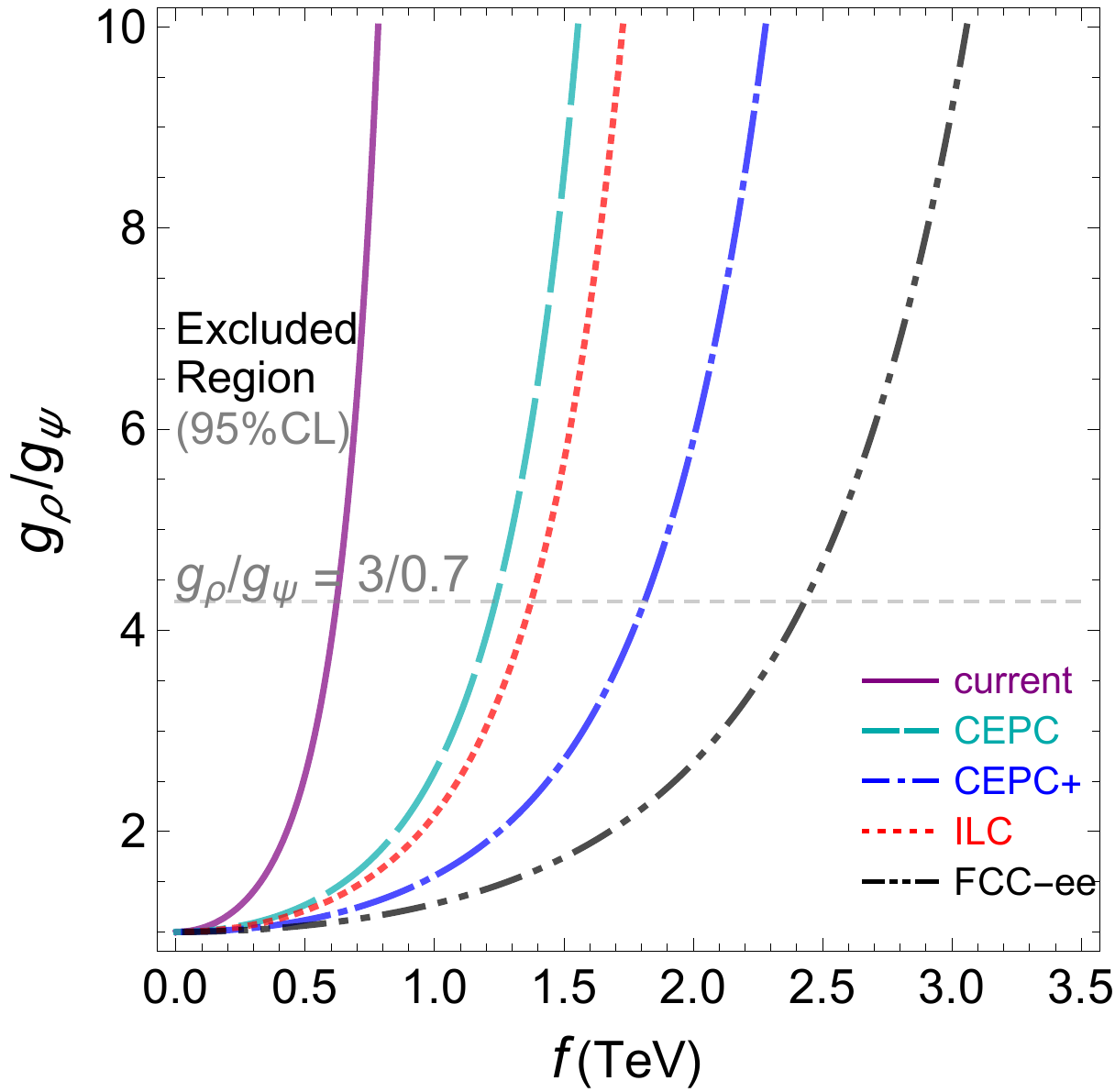}
\caption{ 
Constraints in the $(g_\rho/g_\psi, f)$ plane assuming the only correction to the $Zb\bar{b}$ couplings are given by Eq.~(\ref{eq:ch1}).  Each contour represents the $95\%$ CL constraint and the region in the top-left side of the contour is excluded.  The grey horizontal line corresponds to $g_\rho/g_\psi = 3/0.7$ . In purple is the constraint we
obtain using the present uncertainties on the EWPOs; in cyan, blue, red and black are the
constraints expected with the future measurements at CEPC, CEPC+, ILC and FCC-ee,
respectively.}
\label{fig:ch1}
\end{figure}

Composite Higgs models usually predicts a large correction to the $Zb_L\bar{b}_L$ coupling, since a sizable mixing between the third generation quarks and the strong dynamics is needed to generate the large top mass.  The correction to the $Zb_R\bar{b}_R$ coupling is usually much smaller, unless one specifically extend the fermion sector to generate a large correction ({\it e.g.}, as in Ref.~\cite{DaRold:2010as}).  It was pointed out in Ref.~\cite{Agashe:2006at} that an $O(4)$ symmetry, which is the $SU(2)_L \otimes SU(2)_R$ symmetry, analogous to the custodial symmetry protecting the weak isospin, with the addition of a left-right parity $P_{LR}$, could be used to protect the $Zb_L\bar{b}_L$ coupling, such that a natural composite Higgs model can be consistent with EW precision constraints.  Nevertheless, in realistic models there still exist corrections to the $Zb_L\bar{b}_L$ coupling because 1) the $P_{LR}$ symmetry can only protect the $Zb_L\bar{b}_L$ coupling at zero momentum and 2) there are also several contributions that explicitly break $P_{LR}$.  These corrections could become relevant if the constraint on the $Zb_L\bar{b}_L$ coupling is significantly improved at future $e^+e^-$ colliders.  Ref.~\cite{Grojean:2013qca} estimates the size of different contributions to the $Zb_L\bar{b}_L$ coupling in minimal composite Higgs models with custodial protection.  (Also see Ref.~\cite{Panico:2015jxa} for a recent review.)  While the magnitudes and signs of different contributions are rather model dependent, the leading correction usually comes from $P_{LR}$ breaking effects of fermion loops and is
\begin{equation}
\frac{\dgLb}{g^{\rm SM}_{Lb}} \simeq \frac{y^2_t}{16\pi^2} \frac{v^2}{f^2} \log{\left( \frac{m^2_\rho}{m^2_4} \right)} \,,  \label{eq:ch1}
\end{equation}
where $y_t$ is the top Yukawa coupling, $m_\rho$ is the mass of the $\rho$ meson which cuts off the loop correction, and $m_4$ is the mass of the 4-plet composite quarks which is essentially the mass of the (lightest) top partner up to some corrections from mixing.  Taking Eq.~(\ref{eq:ch1}) as an equality, the results in Table~\ref{tab:ccif} (assuming $\dgRb=0$) can be interpreted in terms of constraints in the $(g_\rho/g_\psi, f)$ plane where $g_\rho \equiv m_\rho/f$ and $g_\psi \equiv m_4/f$. This is shown in Fig.~\ref{fig:ch1}, where each contour represents the $95\%$ CL constraint and the region in the top-left side of the contour is excluded.  The grey horizontal line has $g_\rho/g_\psi = 3/0.7$ which corresponds to the benchmark point $m_\rho = 3$~TeV and $m_4=700$~GeV of Ref.~\cite{Grojean:2013qca}.  

Since $g_\rho/g_\psi$ is typically bounded to be a few times one, future $e^+e^-$ colliders can constrain $f$ to be at least a few TeVs thanks to the measurement of the $Zb\bar{b}$ couplings.  This is comparable to the constraints from the direct searches of top partners at the next run of the LHC, given that the mass of the top partner can not be much larger than $f$ in order to obtain the correct Higgs mass~\cite{Matsedonskyi:2012ym}.  The constraints from the $Zb\bar{b}$ couplings is significantly stronger than the ones from oblique parameters but weaker than the ones from Higgs precision measurement, and in particular from the $HZZ$ vertex, quoted in Ref.~\cite{Fan:2014vta}.  The latter can, in fact, constrain $f$ at the level of $\sim2.8$~TeV (CEPC) and $\sim3.9$~TeV (FCC-ee) at $95\%$CL.  Other studies of future constraints on composite Higgs models can be found in Ref.~\cite{Thamm:2015zwa,Barducci:2015aoa}. To conclude, while the constraints from $Zb\bar{b}$ couplings has a stronger model dependence, it is complementary to the constraints from oblique corrections, Higgs precision measurements and direct searches and can be very helpful for the discrimination of different composite Higgs models.

 %%%%%%%%%%%%%%%%%%%%%%%%%%%%%%%%%%%%%%%%%%%%%%%%%%%%%%%%
 
 \subsection{The Beautiful Mirror Model}

\begin{figure}[t]
\centering
\includegraphics[width=6cm]{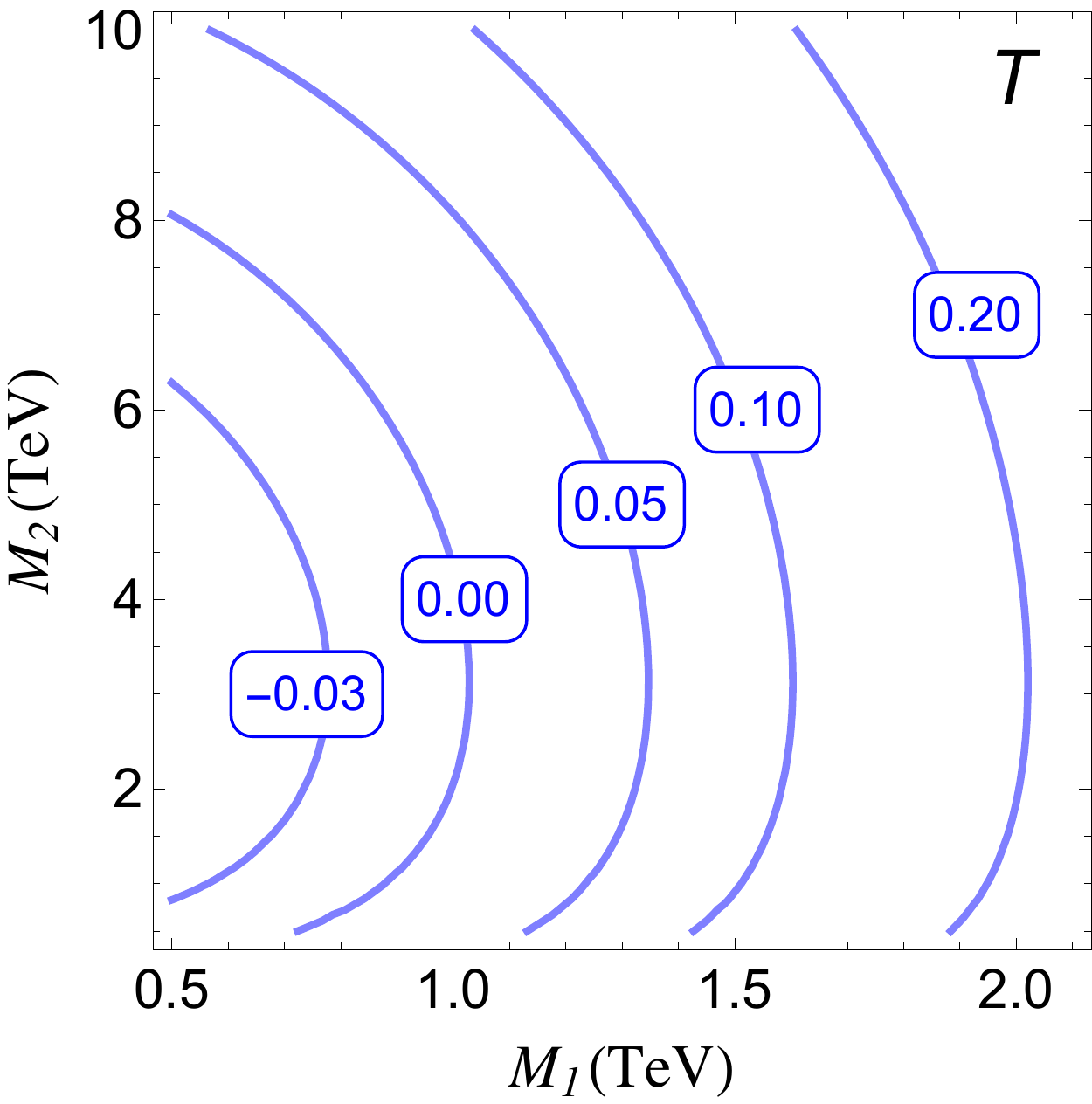} \hspace{0.7cm}
\includegraphics[width=7.8cm]{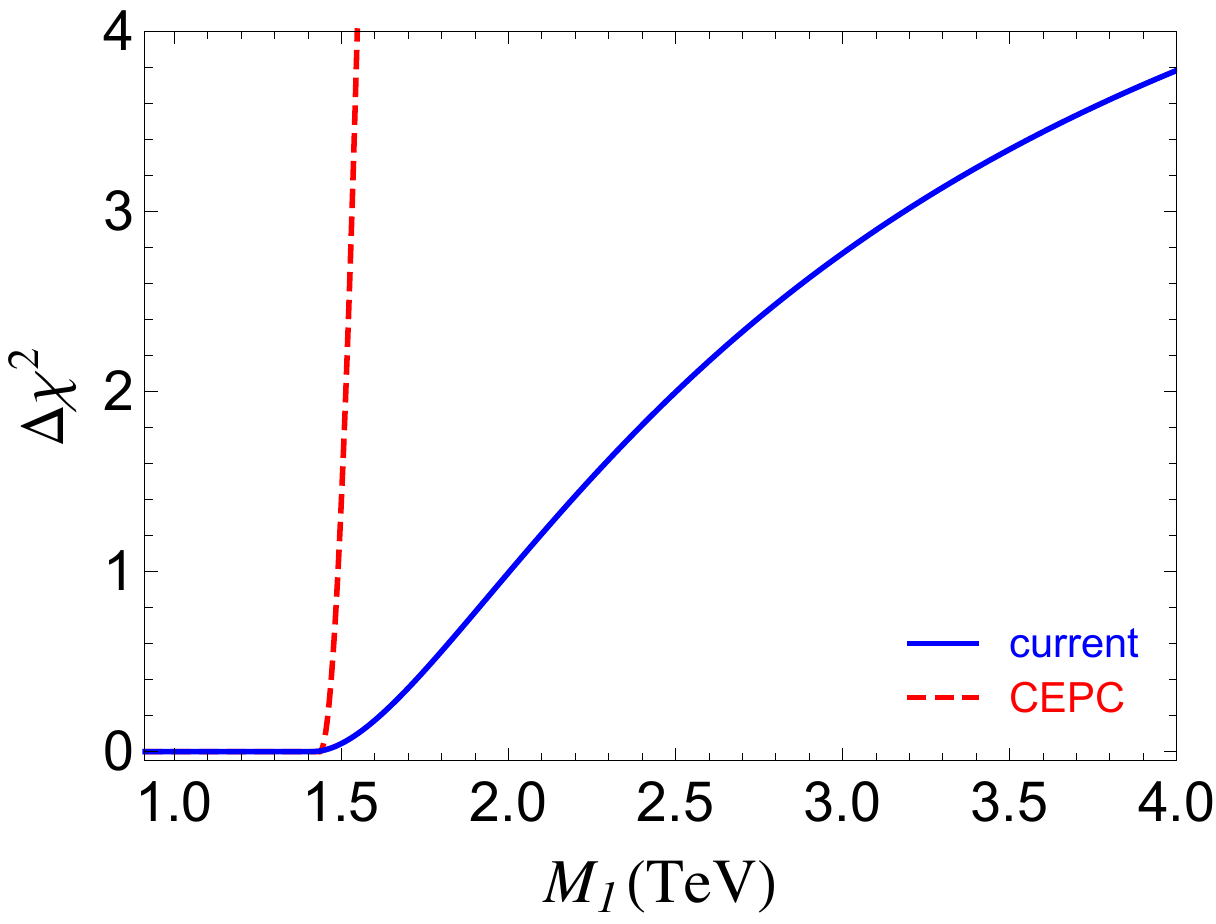} 
\caption{{\bf Left}: The value of $T$ parameter in the $(M_1, M_2)$ plane, while $\dgLb$ and $\dgRb$ are fixed to the current best fit values, $\dgLb = 0.0030$ and $\dgRb=0.0176$.  {\bf Right}: $\Delta\chi^2$ as a function of $M_1$, obtained from the global fit of the EW precision data to the model while marginalizing over other model parameters.  The $\Delta\chi^2$ is measured from the minimum $\chi^2$ that can be reached in this model. The solid blue line is obtained from the current data; the dashed red line is from hypothetical CEPC data, while we na\"ively assume that the central values of $S$, $T$, $\dgLb$, $\dgRb$ are the same as the current ones.  The horizontal axis starts at 912~GeV, the current 95\%~CL bound from the direct searches of vector-like quarks at the 8~TeV LHC.}
\label{fig:BM-1}
\end{figure}

In the Beautiful Mirror Model proposed in Ref.~\cite{Choudhury:2001hs}, the modifications to the $Zb\bar{b}$ couplings are caused by the mixing of the bottom quark and new heavy vector-like quark(s)\footnote{See also~\cite{Aguilar-Saavedra:2013qpa} for a more recent analysis of the fit of the $Zb\bar{b}$ couplings in models with vector-like quarks.}.  In the simplest case with only one bottom-partner $B$, the shifts in the couplings are given by $\dgLb =~(t_3 + 1/2) s^2_L$ and $\dgRb = t_3 s^2_R$, where $t_3$ is the charge of the new bottom quark, $B$, under $SU(2)_L$ and $s_{L(R)}$ is the sine of the left(right)-handed mixing angle between the SM $b$ quark and $B$.  To obtain a shift in $g_{Rb}$, the new quark $B$ can not be a $SU(2)_L$ singlet.  If $B$ has $t_3=-\frac{1}{2}$, $\dgRb$ would be negative and one would need a large shift in the coupling to flip its sign, $g_{Rb}\approx - g^{\rm SM}_{Rb}$, in order to resolve the $\AbFB$ discrepancy.  Such a large shift requires a very light $B$ quark and a large custodial symmetry breaking and has been almost completely probed by LHC direct searches for vector-like quarks~\cite{Morrissey:2003sc,Kumar:2010vx,Batell:2012ca}. A more appealing choice is  $t_3=\frac{1}{2}$, that can lead to a good fit of the EWPOs, without a too light $B$ quark, thanks to a positive contribution to $g_{Rb}$. This scenario was denoted as the ``Top-less Mirror" in the original paper\cite{Choudhury:2001hs}, since there is no top-partner in the new doublet quark.  The new doublet quark alone can not simultaneously generate a sizable enough $\dgLb$, but this can be easily achieved by introducing an additional singlet.  Therefore, we discuss an extension of the SM with the following vector-like quarks,
\begin{align}
\Psi_{L,R} =&~ \bpm B \\ X \epm   \sim (3,2,-5/6) \,, \\
\hat{B}_{L,R}  \sim & ~~  (3,1,-1/3) \,,
\end{align}
where the numbers in the bracket denotes representations under $SU(3)_c$, $SU(2)_L$, and the $U(1)_Y$ hypercharge.  The relevant terms in the Lagrangian are given by
\begin{equation}
-\mathcal{L} \supset M_1 \bar{\Psi}_L \Psi_R + M_2 \bar{\hat{B}}_L \hat{B}_R + y_1 \bar{Q}_L H b_R + y_L \bar{Q}_L H \hat{B}_R + y_R \bar{\Psi}_L \tilde{H} b_R +\mbox{h.c.} \,,
\end{equation}
which leads to the following $3\times 3$ mass matrix $\mathcal{M}_B$ for the bottom-like quarks while the mass of the charge $-4/3$ quark $X$ is simply given by $M_1$,
\begin{equation}
\mathcal{M}_B = \bpm Y_1 & 0  & Y_L \\ Y_R & M_1 & 0 \\ 0 & 0 & M_2 \epm, ~~~~~~\mathcal{M}_X = M_1 \,,
\end{equation}
where $Y_i = \frac{y_i v}{\sqrt{2}}$.  The shifts in the $Zb\bar{b}$ couplings are thus given by
\begin{equation}
\dgLb = \frac{Y^2_L}{2M^2_2}  \,, ~~~~~   \dgRb =  \frac{Y^2_R}{2M^2_1}  \,.  \label{eq:y123}
\end{equation}
The new quarks contributes also to the $T$ parameter through fermion loops.  Ignoring the small bottom mass and the higher order terms, the $T$~parameter is given by
\begin{equation}
T \approx \frac{3}{16\pi^2 \alpha v^2} \left[ \frac{16}{3} \dgRb^2 M^2_1 + 4 \dgLb^2 M^2_2 - 4\dgLb \frac{M^2_2 \, m^2_{\rm top}}{M^2_2-m^2_{\rm top}} \log{\left(\frac{M^2_2}{m^2_{\rm top}}\right)}    \right] \,.  \label{eq:Tsimp}
\end{equation}
If $\dgLb$ and $\dgRb$ are fixed to non-zero values, as preferred by the current EW measurements, the $T$ parameter will increase if one increases the mass scales of the new quarks, as shown in the left plot of Fig.~\ref{fig:BM-1}.  This is because larger Yukawa couplings are needed if one wants to fix $\dgLb$ and $\dgRb$ while increasing $M_1$ and $M_2$, as shown in Eq.~(\ref{eq:y123}).  Since the desired value of $\dgRb$ is much larger than the one of $\dgLb$, the $T$ parameter is more sensitive to $M_1$.  The contribution to the $S$ parameter is very small in this model, and we assume it to be zero for simplicity.  The right plot of Fig.~\ref{fig:BM-1} (solid blue line) shows the value of $\Delta\chi^2$ (measured from the minimum $\chi^2$ that can be reached in this model) as a function of $M_1$, obtained from the global fit of the current EW precision data, while other model parameters are chosen to minimize $\Delta\chi^2$ for a given value of $M_1$.  At small $M_1$ ($\lesssim 1.4$~TeV), one could tune the value of $M_2$ to obtain the best-fit value of the $T$~parameter ($\approx 0.05$ if fixing $S=0$).  For larger values of $M_1$, the agreement with data is significantly worse due to the tension between the $T$~parameter and $\dgRb$.  For comparison, we also show the $\Delta\chi^2$ curve for CEPC (dashed red curve), assuming that the central values of $S$, $T$, $\dgLb$, $\dgRb$ are the same as the current ones: CEPC will tightly constrains $M_1$ to be below $\sim1.6$~TeV.

 An additional upper bound on $M_1$ can be obtained by requiring the theory to be perturbative. In particular, Eq.~(\ref{eq:y123}) implies $y_R \approx 2\sqrt{\delta g_{Rb}} \frac{M_1}{v}$.  Assuming the true values of $\dgLb$ and $\dgRb$ are within $68\%$~CL of the current measurements, we have $\dgRb\gtrsim 0.0075$ and hence
\begin{equation}
y_R \gtrsim \frac{M_1}{1.4~\rm{TeV}} \, .
\end{equation}
Therefore, $M_1$ can not be much larger than a few TeV without the Yukawa coupling, violating perturbativity bounds.

The LHC is directly searching for the mirror quarks.  As pointed out in Ref.~\cite{Kumar:2010vx}, the charge $-4/3$ exotic quark $X$ decays to $b$ and $W$ with the same sign of electric charges, which is extraordinary in theory but hard to capture in experiments, since it is very hard (if not impossible) to measure the charge of $b$-jets.  As such, the strongest bounds on $X$ come from searches of top partners decaying to $b\,W$.  The recent CMS analysis~\cite{CMS:2014dka} sets a lower limit of 912 GeV at the $95\%$ CL for a pair produced top vector-like quark with 100\%BR to $b\,W$, using 8 TeV data. Currently, this is the most stringent constraint on $M_1$.  There also exist bounds from bottom partner searches ({\it e.g.} Ref.~\cite{CMS:2014bfa}) which are slightly weaker.

The current constraint from the LHC is consistent with the one from EW precision data shown in the right plot of Fig.~\ref{fig:BM-1}.  The situation may get much more interesting in the future: the bounds on the mirror quark masses are expected to reach (2-2.5) TeV at the 14~TeV LHC \cite{Kumar:2010vx, Alvarez:2013qwa} using the single production channel.  The HL-LHC with $3000\,{\rm fb}^{-1}$ data could probably  push the bound further to above 3~TeV, if the data is consistent with SM~\cite{crtweb}.  Such a bound would be in tension with the current EW precision data.  If no mirror quark is found during the LHC run, it could be an indication that either 1) the LEP $\AbFB$ discrepancy is not due to NP or it comes from some NP other than the Beautiful Mirror Model\footnote{Another possible solution would be a $Z'$ near $Z$-pole, as proposed in Ref.~\cite{Dermisek:2011xu, Dermisek:2012qx}. Future $e^+e^-$ colliders could perform a much better scan around the $Z$-pole compared with LEP and provide significantly better discriminating power between the SM and this scenario. } or 2) the underlying NP is some extension or modification of the ``minimal" Beautiful Mirror Model which evades the constraints from direct searches, still producing a good fit to EW precision data.

Finally, we comment on the fact that the Beautiful Mirror Model predicts a significant modification to the $hb\bar{b}$ and $hgg$ couplings\cite{Batell:2012ca},
\begin{equation}
\frac{y_{hb\bar{b}}}{(y_{hb\bar{b}})_{\rm{SM}}} \simeq 1-2(\delta g_{Lb}+ \delta g_{Rb})\,, \hspace{1cm}
\frac{g_{hgg}}{(g_{hgg})_{\rm{SM}}} \simeq 1+2(\delta g_{Lb}+ \delta g_{Rb}) \,.
\end{equation}
The current best-fit values of $\dgLb$ and $\dgRb$ predict a $\sim4\%$ deviation on the $hb\bar{b}$ and $hgg$ couplings which will be detectable at a future Higgs Factory~\cite{CEPCPreCDR, Baer:2013cma, Gomez-Ceballos:2013zzn}.

%%%%%%%%%%%%%%%%%%%%%%%%%%%%%%%%%%%%%%%%%%%%%%%%%%%%%%%%
%%%%%%%%%%%%%%%%%%%%%%%%%%%%%%%%%%%%%%%%%%%%%%%%%%%%%%%%

\section{Conclusion}
\label{sec:con}

Precision measurements of SM couplings are the central goal of future lepton colliders. Such measurements are complementary to direct searches at high energy proton colliders. 
In this paper, we have extracted the constraints on possible modifications of the $Zb\bar{b}$ couplings from the SM predictions by performing global fits of both the current precision data and the prospective data from future $e^+e^-$ colliders.  We pointed out that the world average value of the strong coupling constant contains non-trivial information and should be included in the global fit of models with non-zero NP contributions to the $Zb\bar b$ vertex, which has not been pointed out elsewhere.  For the future colliders, we summarized the set of observables most important for improving the $Zb\bar{b}$ coupling constraints and compared the precision reaches at CEPC, ILC and FCC-ee.
We studied both the case that the results are SM-like and the one that the $Zb\bar{b}$ couplings deviate significantly from the SM prediction as suggested by the LEP $\AbFB$ discrepancy.  For the latter case, we showed that even if we assume that the future measurements will point towards smaller values of $\dgLb$ and $\dgRb$ within $68\%$~CL of the current measurements, any one of the proposed $e^+e^-$ colliders will be able to rule out the SM with more than $99.9999\%$~CL, equivalent to 5 standard deviations for a one-dimensional Gaussian distribution.

Finally, we studied the implication on NP models from the improvements of the $Zb\bar{b}$ coupling constraints. 
We first considered generic 2HDMs, in which the limits from precision $Zb\bar{b}$ measurements are complementary to those from the LHC searches of charged Higgs bosons. In particular, future measurements of the $Zb\bar b$ vertex will be able to test Type II 2HDMs at low values of $\tan\beta$ ($\tan\beta\lesssim 1-4$), even in the mass range $160\,{\rm{GeV}}<m_{H^\pm}<180$ GeV, presently not covered by direct LHC searches. We then considered composite Higgs models, where deviations in the $Zb\bar{b}$ couplings are generically expected. Measurement at future lepton colliders can probe composite resonances with masses of multiple TeVs, possibly beyond the reach of the LHC. Finally, in the literature there have been new physics models motivated by the long standing LEP anomaly in $\AbFB$.  As an example, we considered the so called Beautiful Mirror Model in which new fermions mix with the SM bottom quark. We find that, for the minimal scenario, future precision measurements of $Zb\bar{b}$ could put a strong upper bound on the new fermion masses at around 1.6 TeV, if the measured central values of $S$, $T$, $\dgLb$ and $\dgRb$ are the same as the current ones. Future LHC searches of vector-like quarks will be able to fully probe this mass range.

We have also shown that the particular models considered in this paper generically predict deviations in the Higgs couplings, which can also be measured very precisely at the Higgs factory stage of the lepton colliders. The interplay between Higgs and $Z$ precision measurements will be very valuable at future $e^+e^-$ machines, in extracting maximal information about new physics.

%%%%%%%%%%%%%%%%%%%%%%%%%%%%%%%%%%%%%%%%%%%%%%%%%%%%%%%%
%%%%%%%%%%%%%%%%%%%%%%%%%%%%%%%%%%%%%%%%%%%%%%%%%%%%%%%%

\subsection*{Acknowledgments}
We would like to thank Jens Erler for discussion. The research of SG at Perimeter Institute is supported by the Government of Canada through Industry Canada and by the Province of Ontario through the Ministry of Economic Development $\&$ Innovation. SG acknowledge support by the Munich Institute for Astro- and Particle Physics (MIAPP) of the DFG cluster of excellence {\it{Origin and Structure of the Universe}}. JG is supported in part by the Chinese Academy of Science (CAS) International Traveling Award under Grant H95120N1U7 and the CAS Center for Excellence in Particle Physics (CCEPP).  LTW acknowledges the hospitality of the Center for Future High Energy Physics (CFHEP) where part of this work was done.

%%%%%%%%%%%%%%%%%%%%%%%%%%%%%%%%%%%%%%%%%%%%%%%%%%%%%%%%
%%%%%%%%%%%%%%%%%%%%%%%%%%%%%%%%%%%%%%%%%%%%%%%%%%%%%%%%

\appendix

\section{Theoretical uncertainties}
\label{app:theo}

We follow Ref.~\cite{Fan:2014vta} for the treatment of theoretical uncertainties, which we discuss in this Appendix.  
Given a set of model parameters $\boldsymbol{\theta}$, the true value of a particular observable $x$ is predicted %from theory 
to be some certain value, $x_{\rm th}$, up to some uncertainty, $\delta$.  This uncertainty could come from the omission of higher order terms in a fixed order calculation, or the subtlety in the definition of certain quantities ({\it e.g.} the top mass). 
As such, it is strictly speaking not a statistical quantity and there is no good reason to assume it follows a Gaussian distribution.  We assume $x$ takes a probability density function $h(x;\boldsymbol{\theta})$ that is flat within $x_{\rm th} \pm \delta$ and zero elsewhere,
\begin{equation}
h(x;\boldsymbol{\theta}) = \left\{ \begin{matrix}  \frac{1}{2\delta} \mbox{ ~ if  }|x-x_{\rm th}| \leq \delta  \\ 0  \mbox{   ~~~if  }|x-x_{\rm th}| > \delta  \end{matrix}  \right. \,.
\end{equation}
On the other hand, we assume the measured value, $x_{\rm mea}$, takes a Gaussian distribution $g(x_{\rm mea}|x)$ centered around the true value $x$ with standard deviation $\sigma$.

\begin{figure}[h]
\centering
\includegraphics[width=7cm]{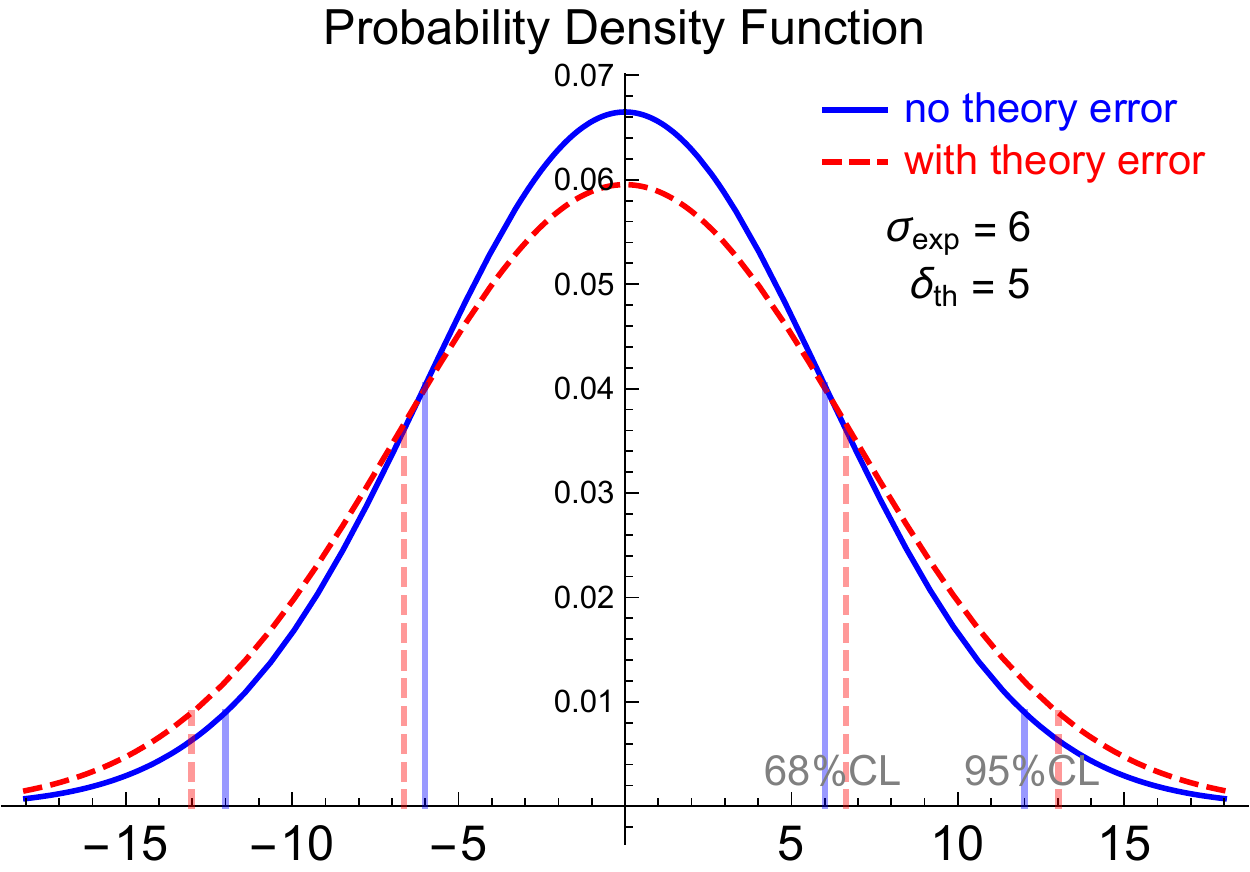}
\caption{Probability Density Functions (p.d.f.) for theoretical uncertainty $\delta_{\rm th} =5$, experimental uncertainty $\sigma_{\rm exp}=6$ and the combined total uncertainty, assuming a zero central value. The blue solid line shows the p.d.f. with the experimental uncertainty only, which follows a Gaussian distribution.  The red dashed line shows the p.d.f. with experimental and theoretical uncertainties combined with Eq.~(\ref{eq:pdff}).  The light vertical lines shows the corresponding $68\%$ and $95\%$\,CL bounds.  Without the theoretical uncertainty, the $68\%$ and $95\%$\,CL bounds are $\pm6$ and $\pm12$; with the theoretical uncertainty, the $68\%$ and $95\%$\,CL bounds are $\pm6.65$ and $\pm13.0$.}
\label{fig:rbth1}
\end{figure}

The distribution of the measure value $x_{\rm mea}$ given a set of model parameters $\boldsymbol{\theta}$ is thus obtained by convolution:
\begin{align}
f(x_{\rm mea};\boldsymbol{\theta}) =&~ \int \! dx ~ g(x_{\rm mea}|x) h(x;\boldsymbol{\theta}) \nonumber\\
=&~ \frac{1}{4\delta} \left(  {\rm erf}\Big(\frac{x_{\rm mea}-x_{\rm th}+\delta}{\sqrt{2}\sigma}\Big) - {\rm erf}\Big(\frac{x_{\rm mea}-x_{\rm th}-\delta}{\sqrt{2}\sigma}\Big)  \right) \,,  \label{eq:pdff}
\end{align}
which reduces to the Gaussian distribution in the limit $\delta\to 0$. Eq.~(\ref{eq:pdff}) can be implemented in a global fit  with $N$ observables with a modified $\chi^2$ function (assuming no correlation)
\begin{equation}
\chi^2_{\rm mod} = \sum^N_{j=1} \left[ -2\log \left( \frac{1}{4\delta_j} \left( {\rm erf}\left(\frac{M_j-O_j+\delta_j}{\sqrt{2}\sigma_j}\right) - {\rm erf}\left(\frac{M_j-O_j-\delta_j}{\sqrt{2}\sigma_j}\right)  \right) \right)  -2\log{(\sqrt{2\pi}\sigma_j)} \right] \,,  \label{eq:therr}
\end{equation}
where for each observable $j$, $M_j$ is the measured value, $O_j$ is the predicted value, $\sigma_j$ is the experimental uncertainty and $\delta_j$ is the theoretical uncertainty.  
$R^0_b$ and $\AbFB$($\Ab$) are directly related to the $Zb\bar{b}$ couplings, and their theoretical uncertainties are most important to our study.  The theoretical uncertainty of $R^0_b$ ($\delta_{\rm th} R^0_b$) is estimated to be $1.5\times10^{-4}$ from two-loop diagrams without closed fermion loops and higher-order contributions \cite{Freitas:2014hra} and could be reduced to a few times $10^{-5}$ assuming the 2-loop and 3-loop computations will be completed in the future \cite{Fan:2014axa}.  Naively, one would expect it to have an impact, especially for FCC-ee which will be able to measure $R^0_b$ to a precision of about $6\times 10^{-5}$.  However, even with a conservative estimation, $\delta_{\rm th} R^0_b = 5\times 10^{-5}$, and for FCC-ee, we find the change of the total uncertainty from the inclusion of the theoretical uncertainty rather small, as the $68\%$\,CL bound changes from $6\times 10^{-5}$ to $6.65\times10^{-5}$.  The corresponding probability density functions are shown in Fig.~\ref{fig:rbth1}, for which we have set the central value to zero and scaled up the uncertainties by $10^5$ for convenience. Given that the estimations of the future experimental uncertainties are still very preliminary, we ignore this small effect due to the theoretical uncertainty of $R^0_b$ in our study.

With (longitudinal) beam polarization, $\Ab$ can be directly measured.  Without beam polarization, $\AbFB$ can be measured, which is related to $\Ab$ by $\AbFB = \frac{3}{4}\Ae \Ab$.  The value of $\Ae$ can be extracted experimentally by the measurement of $\AlFB$.  Therefore, the theoretical uncertainty of $\AbFB$ also only comes from $\Ab$, which can be parameterized by the theoretical uncertainty of $\sin^2{\theta^b_{\rm eff}}$ (The overall form factors of $g_{Lb}$ and $g_{Rb}$ cancel in the ratio.)  $\Ab$ is numerically not very sensitive to $\sin^2{\theta^b_{\rm eff}}$.  At the leading order in the SM, one has
\begin{equation}
\delta_{\rm th} \Ab \approx -0.64 \, \delta_{\rm th} \sin^2{\theta^b_{\rm eff}} \,, \hspace{1cm}  
\delta_{\rm th} \AbFB \approx -0.070 \, \delta_{\rm th} \sin^2{\theta^b_{\rm eff}} \,,
\end{equation}
where $\delta_{\rm th}$ denotes the theoretical uncertainty of the corresponding quantity.  Even with the current theoretical uncertainty of $\sin^2{\theta^b_{\rm eff}}$, which is about $5\times 10^{-5}$ \cite{Awramik:2008gi}, $\delta_{\rm th} \Ab$ and $\delta_{\rm th} \AbFB$ are much smaller than the future experimental precisions and can be safely ignored.

The effects of the theoretical uncertainties of other quantities, such as the top mass and $W$ mass, are important in general ({\it e.g.} for the $S$ and $T$ parameters, as pointed out in Ref.~\cite{Fan:2014vta}) but do not directly affect the $Zb\bar{b}$ couplings.  We implemented these theoretical uncertainties and found that the changes of the $Zb\bar{b}$ coupling constraints are negligible.

%%%%%%%%%%%%%%%%%%%%%%%%%%%%%%%%%%%%%%%%%%%%%%%%%%%%%%%%
%%%%%%%%%%%%%%%%%%%%%%%%%%%%%%%%%%%%%%%%%%%%%%%%%%%%%%%%

%%%%%%%%%%%%%%%%%%%%%%%%%%%%%%%%%%%%%%%%%%%%%%%%%%%%%%%%
%%%%%%%%%%%%%%%%%%%%%%%%%%%%%%%%%%%%%%%%%%%%%%%%%%%%%%%%

\providecommand{\href}[2]{#2}\begingroup\raggedright\endgroup

%%%%%%%%%%%%%%%%%%%%%%%%%%%%%%%%%%%%%%%%%%%%%%%%%%%%%%%%
%%%%%%%%%%%%%%%%%%%%%%%%%%%%%%%%%%%%%%%%%%%%%%%%%%%%%%%%


\begin{thebibliography}{10}

\bibitem{Baer:2013cma}
H.~Baer, T.~Barklow, K.~Fujii, Y.~Gao, A.~Hoang, et~al., {\it {The
  International Linear Collider Technical Design Report - Volume 2: Physics}},
  \href{http://arxiv.org/abs/1306.6352}{{\tt arXiv:1306.6352}}.

\bibitem{Gomez-Ceballos:2013zzn}
{\bf TLEP Design Study Working Group} Collaboration, M.~Bicer et~al., {\it
  {First Look at the Physics Case of TLEP}},  {\em JHEP} {\bf 1401} (2014) 164,
  [\href{http://arxiv.org/abs/1308.6176}{{\tt arXiv:1308.6176}}].

\bibitem{CEPCPreCDR}
M.~Ahmad et~al., ``{CEPC-SPPC Preliminary Conceptual Design Report, Volume I:
  Physics and Detector}.'' \url{http://cepc.ihep.ac.cn/preCDR/volume.html},
  2015.

\bibitem{Baak:2013fwa}
M.~Baak, A.~Blondel, A.~Bodek, R.~Caputo, T.~Corbett, et~al., {\it {Working
  Group Report: Precision Study of Electroweak Interactions}},
  \href{http://arxiv.org/abs/1310.6708}{{\tt arXiv:1310.6708}}.

\bibitem{Fan:2014vta}
J.~Fan, M.~Reece, and L.-T. Wang, {\it {Possible Futures of Electroweak
  Precision: ILC, FCC-ee, and CEPC}},
  \href{http://arxiv.org/abs/1411.1054}{{\tt arXiv:1411.1054}}.

\bibitem{Fan:2014axa}
J.~Fan, M.~Reece, and L.-T. Wang, {\it {Precision Natural SUSY at CEPC, FCC-ee,
  and ILC}},  \href{http://arxiv.org/abs/1412.3107}{{\tt arXiv:1412.3107}}.

\bibitem{Curtin:2014cca}
D.~Curtin, R.~Essig, S.~Gori, and J.~Shelton, {\it {Illuminating Dark Photons
  with High-Energy Colliders}},  {\em JHEP} {\bf 1502} (2015) 157,
  [\href{http://arxiv.org/abs/1412.0018}{{\tt arXiv:1412.0018}}].

\bibitem{Peskin:1991sw}
M.~E. Peskin and T.~Takeuchi, {\it {Estimation of oblique electroweak
  corrections}},  {\em Phys.Rev.} {\bf D46} (1992) 381--409.

\bibitem{Peccei:1990uv}
R.~Peccei, S.~Peris, and X.~Zhang, {\it {Nonstandard couplings of the top quark
  and precision measurements of the electroweak theory}},  {\em Nucl.Phys.}
  {\bf B349} (1991) 305--322.

\bibitem{Agashe:2006at}
K.~Agashe, R.~Contino, L.~Da~Rold, and A.~Pomarol, {\it {A Custodial symmetry
  for Zb anti-b}},  {\em Phys.Lett.} {\bf B641} (2006) 62--66,
  [\href{http://arxiv.org/abs/hep-ph/0605341}{{\tt hep-ph/0605341}}].

\bibitem{Haber:1999zh}
H.~E. Haber and H.~E. Logan, {\it {Radiative corrections to the Z b anti-b
  vertex and constraints on extended Higgs sectors}},  {\em Phys.Rev.} {\bf
  D62} (2000) 015011, [\href{http://arxiv.org/abs/hep-ph/9909335}{{\tt
  hep-ph/9909335}}].

\bibitem{ALEPH:2005ab}
{\bf ALEPH, DELPHI, L3, OPAL, SLD, LEP Electroweak Working Group, SLD
  Electroweak Group, SLD Heavy Flavour Group} Collaboration, S.~Schael et~al.,
  {\it {Precision electroweak measurements on the $Z$ resonance}},  {\em
  Phys.Rept.} {\bf 427} (2006) 257--454,
  [\href{http://arxiv.org/abs/hep-ex/0509008}{{\tt hep-ex/0509008}}].

\bibitem{Freitas:2012sy}
A.~Freitas and Y.-C. Huang, {\it {Electroweak two-loop corrections to $sin^{2}
  \theta^{b \bar b}_{eff}$ and $R_{b}$ using numerical Mellin-Barnes
  integrals}},  {\em JHEP} {\bf 1208} (2012) 050,
  [\href{http://arxiv.org/abs/1205.0299}{{\tt arXiv:1205.0299}}].

\bibitem{Baak:2014ora}
{\bf Gfitter Group} Collaboration, M.~Baak et~al., {\it {The global electroweak
  fit at NNLO and prospects for the LHC and ILC}},  {\em Eur.Phys.J.} {\bf C74}
  (2014) 3046, [\href{http://arxiv.org/abs/1407.3792}{{\tt arXiv:1407.3792}}].

\bibitem{Freitas:2014hra}
A.~Freitas, {\it {Higher-order electroweak corrections to the partial widths
  and branching ratios of the Z boson}},  {\em JHEP} {\bf 1404} (2014) 070,
  [\href{http://arxiv.org/abs/1401.2447}{{\tt arXiv:1401.2447}}].

\bibitem{Batell:2012ca}
B.~Batell, S.~Gori, and L.-T. Wang, {\it {Higgs Couplings and Precision
  Electroweak Data}},  {\em JHEP} {\bf 1301} (2013) 139,
  [\href{http://arxiv.org/abs/1209.6382}{{\tt arXiv:1209.6382}}].

\bibitem{Ciuchini:2013pca}
M.~Ciuchini, E.~Franco, S.~Mishima, and L.~Silvestrini, {\it {Electroweak
  Precision Observables, New Physics and the Nature of a 126 GeV Higgs Boson}},
   {\em JHEP} {\bf 1308} (2013) 106,
  [\href{http://arxiv.org/abs/1306.4644}{{\tt arXiv:1306.4644}}].

\bibitem{Flacher:2008zq}
H.~Flacher, M.~Goebel, J.~Haller, A.~Hocker, K.~Monig, et~al., {\it {Revisiting
  the Global Electroweak Fit of the Standard Model and Beyond with Gfitter}},
  {\em Eur.Phys.J.} {\bf C60} (2009) 543--583,
  [\href{http://arxiv.org/abs/0811.0009}{{\tt arXiv:0811.0009}}].

\bibitem{Pomarol:2013zra}
A.~Pomarol and F.~Riva, {\it {Towards the Ultimate SM Fit to Close in on Higgs
  Physics}},  {\em JHEP} {\bf 01} (2014) 151,
  [\href{http://arxiv.org/abs/1308.2803}{{\tt arXiv:1308.2803}}].

\bibitem{Ellis:2014jta}
J.~Ellis, V.~Sanz, and T.~You, {\it {The Effective Standard Model after LHC Run
  I}},  {\em JHEP} {\bf 03} (2015) 157,
  [\href{http://arxiv.org/abs/1410.7703}{{\tt arXiv:1410.7703}}].

\bibitem{Falkowski:2014tna}
A.~Falkowski and F.~Riva, {\it {Model-independent precision constraints on
  dimension-6 operators}},  {\em JHEP} {\bf 02} (2015) 039,
  [\href{http://arxiv.org/abs/1411.0669}{{\tt arXiv:1411.0669}}].

\bibitem{Berthier:2015oma}
L.~Berthier and M.~Trott, {\it {Towards consistent Electroweak Precision Data
  constraints in the SMEFT}},  {\em JHEP} {\bf 05} (2015) 024,
  [\href{http://arxiv.org/abs/1502.02570}{{\tt arXiv:1502.02570}}].

\bibitem{Berthier:2015gja}
L.~Berthier and M.~Trott, {\it {Consistent constraints on the Standard Model
  Effective Field Theory}},  \href{http://arxiv.org/abs/1508.05060}{{\tt
  arXiv:1508.05060}}.

\bibitem{Agashe:2014kda}
{\bf Particle Data Group} Collaboration, K.~Olive et~al., {\it {Review of
  Particle Physics}},  {\em Chin.Phys.} {\bf C38} (2014) 090001.

\bibitem{ATLAS:2014wva}
{\bf ATLAS, CDF, CMS, D0} Collaboration, {\it {First combination of Tevatron
  and LHC measurements of the top-quark mass}},
  \href{http://arxiv.org/abs/1403.4427}{{\tt arXiv:1403.4427}}.

\bibitem{Aad:2014aba}
{\bf ATLAS} Collaboration, G.~Aad et~al., {\it {Measurement of the Higgs boson
  mass from the $H\rightarrow \gamma\gamma$ and $H \rightarrow ZZ^{*}
  \rightarrow 4\ell$ channels with the ATLAS detector using 25 fb$^{-1}$ of
  $pp$ collision data}},  {\em Phys.Rev.} {\bf D90} (2014), no.~5 052004,
  [\href{http://arxiv.org/abs/1406.3827}{{\tt arXiv:1406.3827}}].

\bibitem{CMS:2014ega}
{\bf CMS} Collaboration, {\it {Precise determination of the
  mass of the Higgs boson and studies of the compatibility of its couplings
  with the standard model}}, {\tt CMS-PAS-HIG-14-009}.

\bibitem{Davier:2010nc}
M.~Davier, A.~Hoecker, B.~Malaescu, and Z.~Zhang, {\it {Reevaluation of the
  Hadronic Contributions to the Muon g-2 and to alpha(MZ)}},  {\em Eur.Phys.J.}
  {\bf C71} (2011) 1515, [\href{http://arxiv.org/abs/1010.4180}{{\tt
  arXiv:1010.4180}}].

\bibitem{Murphy:2015cha}
C.~W. Murphy, {\it {Bottom-Quark Forward-Backward and Charge Asymmetries at
  Hadron Colliders}},  \href{http://arxiv.org/abs/1504.02493}{{\tt
  arXiv:1504.02493}}.

\bibitem{Pich:2013sqa}
A.~Pich, {\it {Review of $\alpha_s$ determinations}},  {\em PoS} {\bf
  ConfinementX} (2012) 022, [\href{http://arxiv.org/abs/1303.2262}{{\tt
  arXiv:1303.2262}}].

\bibitem{Ciuchini:2014dea}
M.~Ciuchini, E.~Franco, S.~Mishima, M.~Pierini, L.~Reina, et~al., {\it {Update
  of the electroweak precision fit, interplay with Higgs-boson signal strengths
  and model-independent constraints on new physics}},
  \href{http://arxiv.org/abs/1410.6940}{{\tt arXiv:1410.6940}}.

\bibitem{Choudhury:2001hs}
D.~Choudhury, T.~M. Tait, and C.~Wagner, {\it {Beautiful mirrors and precision
  electroweak data}},  {\em Phys.Rev.} {\bf D65} (2002) 053002,
  [\href{http://arxiv.org/abs/hep-ph/0109097}{{\tt hep-ph/0109097}}].

\bibitem{Aad:2014kga}
{\bf ATLAS} Collaboration, G.~Aad et~al., {\it {Search for charged Higgs bosons
  decaying via $H^{\pm} \rightarrow \tau^{\pm}\nu$ in fully hadronic final
  states using $pp$ collision data at $\sqrt{s} = 8$ TeV with the ATLAS
  detector}},  {\em JHEP} {\bf 1503} (2015) 088,
  [\href{http://arxiv.org/abs/1412.6663}{{\tt arXiv:1412.6663}}].

\bibitem{CMS:2014pea}
{\bf CMS} Collaboration, {\it {Search for a heavy charged
  Higgs boson in proton-proton collisions at sqrts=8 TeV with the CMS
  detector}}, {\tt CMS-PAS-HIG-13-026}.

\bibitem{Craig:2015jba}
N.~Craig, F.~DíEramo, P.~Draper, S.~Thomas, and H.~Zhang, {\it {The Hunt for
  the Rest of the Higgs Bosons}},  {\em JHEP} {\bf 1506} (2015) 137,
  [\href{http://arxiv.org/abs/1504.04630}{{\tt arXiv:1504.04630}}].

\bibitem{DaRold:2010as}
L.~Da~Rold, {\it {Solving the $A_{FB}^b$ anomaly in natural composite models}},
   {\em JHEP} {\bf 1102} (2011) 034,
  [\href{http://arxiv.org/abs/1009.2392}{{\tt arXiv:1009.2392}}].

\bibitem{Grojean:2013qca}
C.~Grojean, O.~Matsedonskyi, and G.~Panico, {\it {Light top partners and
  precision physics}},  {\em JHEP} {\bf 1310} (2013) 160,
  [\href{http://arxiv.org/abs/1306.4655}{{\tt arXiv:1306.4655}}].

\bibitem{Panico:2015jxa}
G.~Panico and A.~Wulzer, {\it {The Composite Nambu-Goldstone Higgs}},
  \href{http://arxiv.org/abs/1506.01961}{{\tt arXiv:1506.01961}}.

\bibitem{Matsedonskyi:2012ym}
O.~Matsedonskyi, G.~Panico, and A.~Wulzer, {\it {Light Top Partners for a Light
  Composite Higgs}},  {\em JHEP} {\bf 01} (2013) 164,
  [\href{http://arxiv.org/abs/1204.6333}{{\tt arXiv:1204.6333}}].

\bibitem{Thamm:2015zwa}
A.~Thamm, R.~Torre, and A.~Wulzer, {\it {Future tests of Higgs compositeness:
  direct vs indirect}},  \href{http://arxiv.org/abs/1502.01701}{{\tt
  arXiv:1502.01701}}.

\bibitem{Barducci:2015aoa}
D.~Barducci, S.~De~Curtis, S.~Moretti, and G.~M. Pruna, {\it {Top pair
  production at a future $e^+e^-$ machine in a composite Higgs scenario}},
  \href{http://arxiv.org/abs/1504.05407}{{\tt arXiv:1504.05407}}.

\bibitem{Aguilar-Saavedra:2013qpa}
J.~Aguilar-Saavedra, R.~Benbrik, S.~Heinemeyer, and M.~Pérez-Victoria, {\it
  {Handbook of vectorlike quarks: Mixing and single production}},  {\em
  Phys.Rev.} {\bf D88} (2013), no.~9 094010,
  [\href{http://arxiv.org/abs/1306.0572}{{\tt arXiv:1306.0572}}].

\bibitem{Morrissey:2003sc}
D.~E. Morrissey and C.~E.~M. Wagner, {\it {Beautiful mirrors, unification of
  couplings and collider phenomenology}},  {\em Phys. Rev.} {\bf D69} (2004)
  053001, [\href{http://arxiv.org/abs/hep-ph/0308001}{{\tt hep-ph/0308001}}].

\bibitem{Kumar:2010vx}
K.~Kumar, W.~Shepherd, T.~M. Tait, and R.~Vega-Morales, {\it {Beautiful Mirrors
  at the LHC}},  {\em JHEP} {\bf 1008} (2010) 052,
  [\href{http://arxiv.org/abs/1004.4895}{{\tt arXiv:1004.4895}}].

\bibitem{CMS:2014dka}
{\bf CMS} Collaboration, {\it {Search for vector-like quarks
  in final states with a single lepton and jets in pp collisions at sqrt s = 8
  TeV}}, {\tt CMS-PAS-B2G-12-017}.

\bibitem{CMS:2014bfa}
{\bf CMS} Collaboration, {\it {Search for pair-produced
  vector-like quarks of charge -1/3 decaying to bH using boosted Higgs
  jet-tagging in pp collisions at sqrt(s) = 8 TeV}}, {\tt CMS-PAS-B2G-14-001}.

\bibitem{Alvarez:2013qwa}
E.~Álvarez, L.~Da~Rold, and J.~I. Sanchez~Vietto, {\it {Single production of an
  exotic bottom partner at LHC}},  {\em JHEP} {\bf 1402} (2014) 010,
  [\href{http://arxiv.org/abs/1311.2077}{{\tt arXiv:1311.2077}}].

\bibitem{crtweb}
G.~Salam and A.~Weiler, ``{Collider Reach ($\beta$)}.''
  \url{http://collider-reach.web.cern.ch/collider-reach/}.

\bibitem{Dermisek:2011xu}
R.~Dermisek, S.-G. Kim, and A.~Raval, {\it {New Vector Boson Near the Z-pole
  and the Puzzle in Precision Electroweak Data}},  {\em Phys.Rev.} {\bf D84}
  (2011) 035006, [\href{http://arxiv.org/abs/1105.0773}{{\tt
  arXiv:1105.0773}}].

\bibitem{Dermisek:2012qx}
R.~Dermisek, S.-G. Kim, and A.~Raval, {\it {Z' near the Z-pole}},  {\em
  Phys.Rev.} {\bf D85} (2012) 075022,
  [\href{http://arxiv.org/abs/1201.0315}{{\tt arXiv:1201.0315}}].

\bibitem{Awramik:2008gi}
M.~Awramik, M.~Czakon, A.~Freitas, and B.~A. Kniehl, {\it {Two-loop electroweak
  fermionic corrections to $\sin^2{\theta^{b\bar{b}}_{\rm eff}}$}},  {\em Nucl.
  Phys.} {\bf B813} (2009) 174--187,
  [\href{http://arxiv.org/abs/0811.1364}{{\tt arXiv:0811.1364}}].

\end{thebibliography}
\end{document}